\documentclass[aip,jcp,preprint]{revtex4-2}
\usepackage[latin2]{inputenc}
\usepackage[english]{babel}
\usepackage{t1enc}
\usepackage{amsfonts}
\usepackage{graphicx}
\usepackage{amssymb}
\usepackage{amsmath}
\usepackage{textcomp}
\usepackage{morefloats}
\usepackage{gensymb}
\usepackage{fullpage}
\usepackage{enumerate}
\usepackage{rotating}
\usepackage{acronym}
\usepackage{scalefnt}
\usepackage{epsfig}
\usepackage{enumitem}
\usepackage{wrapfig}
\usepackage{esint}
\usepackage{float}
\usepackage{upgreek}
\usepackage{bm}
\usepackage{amsbsy}
\usepackage{color}

\makeatletter
\providecommand\barcirc{\mathpalette\@barred\circ}
\def\@barred#1#2{\ooalign{\hfil$#1-$\hfil\cr\hfil$#1#2$\hfil\cr}}

\makeatother

\renewcommand\footnotemark{}

\begin{document}

\title{Practical guide to the statistical mechanics of molecular polaritons}

\author{Csaba F\'abri}
\email{ficsaba@staff.elte.hu}
\affiliation{MTA-ELTE Complex Chemical Systems Research Group, P.O. Box 32, H-1518 Budapest 112, Hungary}

\begin{abstract}
A theoretical approach aimed at the quantum statistical mechanics of a molecular ensemble coupled to a lossless cavity mode is presented.
A canonical ensemble is considered and an approximate formula is devised for the Helmholtz free energy correction due to cavity-molecule coupling,
which enables the derivation of experimentally measurable thermodynamic quantities. The frequency of the cavity mode is assumed to lie in the infrared range.
Therefore, the cavity couples to molecular vibrations and our treatment is restricted to the electronic ground state of the molecule.
The method is tested for an analytically solvable model system of one-dimensional harmonic oscillators coupled to the cavity.
The performance of the approximation and its range of validity are discussed in detail. It is shown that the leading-order correction to the
Helmholtz free energy is proportional to the square of the collective coupling strength. We also demonstrate that the cavity mode does not have
a significant impact on the thermodynamic properties of the system in the collective ultrastrong coupling regime
(the collective coupling strength is comparable to the frequency of the cavity mode).
\end{abstract}

\maketitle

\section{Introduction}

Polaritonic chemistry,\cite{12HuScGe,15GeShHu,15ShGeHu,16Ebbesen,16ChNiBe} a novel and emerging field of physical chemistry, offers new ways
to control the properties of matter and the course of physicochemical processes. It was first demonstrated by Ebbesen and coworkers in a groundbreaking
experiment that the rate of chemical reactions can be altered by strongly coupling molecules to vacuum fields in an optical cavity.\cite{12HuScGe}
Later, it has been shown that strong coupling between molecules and quantised light modes confined in a cavity has a significant impact on the
dynamical,\cite{15ScGeTi,16KoBeMu,17LuFeTo,17HaScSc,18HaScSc,18FrGrCo,18MaDuRi,18DuMaRi,18Vendrell,18RoWeRu,19ReSoGe,19KaAsTo,19GrClFe,19UlGoVe,19CsViHa,19CsKoHa,19TrSa,20GuMu,20GuMu_2,20MaMoHu,20TaMaZh,20FeFrSc,20UlVe,20UlGoVE,20ScRuRo,21SiScRu,21HaScRo,21SoReMi,20SiPiGa,20LiSuNi,20LiChNi,20DaKo,20Spano,21TrSa,21GuKo,21LiNiSu,21TiFeGr,21CeKu,21Cederbaum,21SzBaHa,22SiFe,22Cederbaum,22LiNiHa,22CsVeHa,22FaHaVi,23FaHaCe,23PeKoSt,23NaVe,23ScKo,23Szidarovszky,23Szidarovszky2}
spectroscopic \cite{15GaGaFe,17HeSp,18HeSp,18XiRiDu,18SzHaCs,20SzHaVi,20FaLaHa,21FaHaCe}
and topological \cite{21FaMaHu,22BaUmFa}
properties of molecules as well as on 
chemical reaction rates, \cite{16GaGaFe,16HeSp,19GaClGa,19MaHu,20MaKrHu,20LiNiSu,20HoLaRu,21LiMaHu,21LiMaHu_2,22Schafer,22ScFlRo,22PaCaMa,22DuYu,22SuVe}
also covered by an array of review articles.\cite{17KoMu,18FeGaGa,18RiMaDu,20HeOw,22LiCuSu,22FrGaFe,22SaFeFe}
Common to these observations is the emergence hybrid light-matter states, called polaritonic states which are formed due to strong cavity-molecule coupling and
possess both excitonic and photonic properties.
In addition, well-established methods of electronic structure theory have been extended to describe cavity-molecule interactions.\cite{14RuFlPe,17FlRuAp,20HaRoKj,21ScBuPe,22ScJo}
From a physics perspective, polaritonic chemistry can be regarded as quantum optics with molecules, which,
owing to the internal (electronic, vibrational and rotational) molecular degrees of freedom, clearly requires theoretical approaches that go beyond
standard quantum-optical models dealing with simple two-state quantum emitters.\cite{63JaCu,68TaCu,73HeLi}

Although it was verified that single-molecule strong coupling can be realised,\cite{16ChNiBe} experiments in polaritonic chemistry tipically involve
an ensemble of molecules.\cite{12HuScGe,15GeShHu,15ShGeHu,16Ebbesen} Therefore, we focus on the theoretical description of a molecular ensemble
coupled to a cavity mode and examine such systems from a statistical-mechanical perspective.
More precisely, we assume that the molecular ensemble interacts with a single mode of a lossless infrared cavity.
Since the cavity frequency lies in the infrared range, the cavity mode couples to molecular vibrations (and rotations).
Our treatment is restricted to the electronic ground state and polarization effects (see for example Ref. \citenum{23Szidarovszky}) are neglected.
Using the toolkit of quantum statistical mechanics for a canonical ensemble (constant temperature, volume and number of particles),
the canonical density operator is approximated by perturbative expansion in terms of the cavity-molecule interaction.
Our approach, reminiscent of the cumulant expression used in the statistical mechanics of interacting systems,\cite{07Kardar} takes
collective effects into account in a proper way. The resulting formula for the Helmholtz free energy correction ($\Delta F$) due to
cavity-molecule coupling requires only parameters of the cavity and properties of an isolated molecule. It is also shown that the leading-order
term in $\Delta F$ is proportional to the square of the collective coupling strength $G = g \sqrt{N}$ ($g$ equals the coupling strength parameter
and $N$ denotes the number of molecules in the ensemble).

The limitations of the method presented are as follows.
Although intermolecular interactions become relevant at higher particle densities, we opt to neglect intermolecular forces in the current work.
Molecules are assumed to have the same orientation with respect to the cavity field and a single cavity mode is employed with a homogeneous electric field.
In addition, consequences of the Pauli principle for identical molecules \cite{23Szidarovszky2} are currently disregarded. We believe that these limitations are mainly
of technical nature and they can be overcome without abandoning the core ideas of our approach, which is left for future work. The method is tested for a model system of
one-dimensional harmonic oscillators which is amenable to an analytical solution, allowing for a straightforward comparison of our approximate results to
their exact counterparts in this particular case. Similar studies have been carried out by Rabl and coworkers in Ref. \onlinecite{20PiBeRa} where the thermodynamic 
properties of an ensemble of two-level dipoles coupled to a cavity mode have been investigated.
Our findings corroborate the conclusions drawn in Ref. \onlinecite{20PiBeRa}. We also demonstrate that the current method provides accurate results in the realm of
collective ultrastrong coupling (cUSC) where the collective coupling strength is comparable to the cavity frequency.\cite{20PiBeRa}
Moreover, the cavity mode is shown to have a minor impact on the thermodynamic properties of the system in the cUSC regime.
It is also justified that our approach is in principle applicable to molecules described with (ro)vibrational models of arbitrary complexity,
ranging from standard harmonic-oscillator and rigid-rotor models to sophisticated variational nuclear motion techniques.\cite{12CsFaSz}

\clearpage

\section{Theory}
\label{sec:theory}

An ensemble of $N$ molecules coupled to a single lossless cavity mode is described by the Hamiltonian \cite{04CoDuGr}
\begin{equation}
    \hat{H} = \sum_{i=1}^N \hat{H}^{(0)}_i + \frac{1}{2} \left( \hat{p}_\textrm{c}^2 + \omega_\textrm{c}^2 q_\textrm{c}^2 \right) -
    	g \sqrt{\frac{2 \omega_\textrm{c}}{\hbar}} q_\textrm{c} \mu + \frac{g^2}{\hbar \omega_\textrm{c}} \mu^2
\label{eq:H}
\end{equation}
where $\hat{H}^{(0)}_i$ is the Hamiltonian of the $i$th molecule, $q_\textrm{c}$ and $\hat{p}_\textrm{c}$ denote
the coordinate and momentum associated with the cavity mode, respectively, and $\omega_\textrm{c}$ is the cavity angular frequency.
The cavity-molecule coupling is characterised by the coupling strength parameter $g = \sqrt{\frac{\hbar \omega_\textrm{c}}{2 \epsilon_0 V}}$
where $\epsilon_0$ and $V$ refer to the permittivity and quantisation volume of the cavity, respectively. In addition, $\mu = \vec{\mu} \vec{e}$ is the component 
of the total molecular electric dipole moment $\vec{\mu} =  \sum_{i=1}^N \vec{\mu}_i$ along the field polarization vector $\vec{e}$.
The last term of $\hat{H}$ corresponds to the dipole self energy (DSE) whose relevance has been investigated thoroughly.\cite{18RoWeRu,20ScRuRo,20MaMoHu,22FrGaFe}
It is worth noting that the DSE term contains the square of the dipole moment operator $\mu$. If a single molecule is considered,
the expected value $\langle \Psi_0 | \mu^2 | \Psi_0 \rangle$ over the electronic ground state $|\Psi_0\rangle$ can be written as 
$\langle \Psi_0 | \mu^2 | \Psi_0 \rangle = \sum_i \langle \Psi_0 | \mu | \Psi_i \rangle \langle \Psi_i | \mu | \Psi_0 \rangle $ where
a resolution of identity is inserted in between the two $\mu$ operators ($| \Psi_i \rangle$ denote electronic states). If only the electronic ground state
is taken into account, one can make the approximation $\langle \Psi_0 | \mu^2 | \Psi_0 \rangle \approx |\langle \Psi_0 | \mu | \Psi_0 \rangle|^2$ and
the expected value of $\mu^2$ is replaced by the squared expected value of $\mu$. Similar considerations can be made for the case of $N$ molecules, yielding
$\langle \Psi_0 | \mu^2 | \Psi_0 \rangle \approx \sum_{i} |\langle \Psi^{(i)}_0 | \mu_i | \Psi^{(i)}_0 \rangle|^2 +
\sum_{i \ne j} \langle \Psi^{(i)}_0 | \mu_i | \Psi^{(i)}_0 \rangle \langle \Psi^{(j)}_0 | \mu_j | \Psi^{(j)}_0 \rangle $ with $i,j = 1, \dots, N$
(regarding the validity of this approximation see also Ref. \onlinecite{20LiNiSu} which presented an equilibrium statistical-mechanical treatment
of chemical reactions in cavity).
This expression, used in the present work, also reveals that the DSE term induces intermolecular dipole-dipole interactions
(see also Ref. \onlinecite{23SiScOb}) which are included in our model.

It is useful to partition $\hat{H}$ as
\begin{equation}
    \hat{H} = \hat{H}_0 + \Delta\hat{H}
\end{equation}
where
\begin{equation}
    \hat{H}_0 = \sum_{i=1}^N \hat{H}^{(0)}_i + \hat{H}_\textrm{c}
\label{eq:H0}
\end{equation}
with $\hat{H}_\textrm{c} = \frac{1}{2} \left( \hat{p}^2_\textrm{c} + \omega_\textrm{c}^2 q^2_\textrm{c} \right)$ and
\begin{equation}
    \Delta\hat{H} = -g \sqrt{\frac{2 \omega_\textrm{c}}{\hbar}} q_\textrm{c} \mu + \frac{g^2}{\hbar \omega_\textrm{c}} \mu^2.
\label{eq:deltaH}
\end{equation}
We stress that terms depending on $\mu$ appear solely in $\Delta\hat{H}$.

Molecules in the ensemble are deemed identical and we assume that the time-independent molecular Schr\"odinger equation has been solved,
\begin{equation}
    \hat{H}^{(0)}_i|\alpha_i\rangle = \epsilon_{\alpha_i} |\alpha_i\rangle
\end{equation}
where $\epsilon_{\alpha_i}$ and $|\alpha_i\rangle$ are energy levels and eigenstates of the $i$th molecule, respectively.
Similarly, one has to consider the time-independent Schr\"odinger equation for the cavity mode, that is,
\begin{equation}
    \hat{H}_\textrm{c} |n\rangle = \epsilon_n |n\rangle
\end{equation}
where $|n\rangle$ are Fock states of the cavity mode (with $n=0,1,2,\dots$) and the corresponding energy levels $\epsilon_n$ read
\begin{equation}
     \epsilon_n = \hbar \omega_\textrm{c} \left( n+\frac{1}{2} \right).
\end{equation}
Since $\hat{H}_0$ of Eq. \eqref{eq:H0} equals the sum of uncoupled Hamiltonians, energy levels and eigenfunctions of $\hat{H}_0$ can be expressed as
\begin{equation}
    \hat{H}_0 |\alpha_1\rangle \dots |\alpha_N\rangle |n\rangle = \left( \sum_{i=1}^N \epsilon_{\alpha_i}+\epsilon_n \right) |\alpha_1\rangle \dots |\alpha_N\rangle |n\rangle
\end{equation}
where the eigenfunctions $|\alpha_i\rangle$ can be combined into $|\alpha \rangle = \otimes_{i=1}^N | \alpha_i \rangle$.

As is well known from statistical mechanics,\cite{07Kardar} the thermodynamic quantity central to canonical ensembles is
the Helmholtz free energy $F$ which can be obtained by the relation
\begin{equation}
    F = -k_\textrm{B} T \ln Q = -\frac{1}{\beta} \ln Q
\end{equation}
where $Q = \textrm{tr}[\exp(-\beta \hat{H})]$ is the canonical partition function, $T$ is the temperature,
$k_\textrm{B}$ denotes the Boltzmann constant and $\beta = 1/(k_\textrm{B}T)$.
With this, one can readily determine the free energy correction due to the cavity-molecule coupling,
\begin{equation}
    \Delta F = F - F_0 = -\frac{1}{\beta} \ln \frac{Q}{Q_0} = -\frac{1}{\beta} \ln \frac{\textrm{tr}[\exp(-\beta \hat{H})]}{\textrm{tr}[\exp(-\beta \hat{H}_0)]}
\label{eq:deltaF}
\end{equation}
where $F$ and $F_0$ are free energies of the coupled ($\hat{H}$) and uncoupled ($\hat{H}_0$) cavity-molecule ensembles.
$Q_0 = \textrm{tr}[\exp(-\beta \hat{H}_0)]$ can be expressed as the product of cavity and
molecular partition functions (see later paragraphs for more information).

The major difficulty of evaluating $ \Delta F$ obviously stems from the fact that $\hat{H}$ includes the cavity-molecule interaction Hamiltonian
$\Delta \hat{H}$ which, according to Eq. \eqref{eq:deltaH}, couples the different molecular and cavity degrees of freedom.
This problem is circumvented by the Taylor expansion
\begin{equation}
    \exp(-\beta \hat{H}) = \sum_{k=0}^\infty \frac{(-\beta)^k}{k!} (\hat{H}_0+\Delta \hat{H})^k = \exp(-\beta \hat{H}_0) + \sum_{k=1}^\infty \hat{\Delta}_k
\label{eq:expansion}
\end{equation}
where $\hat{\Delta}_k$ collects every possible expansion term which includes $\Delta \hat{H}$ exactly $k$ times.
Taking into account that $\hat{H}_0$ and $\Delta \hat{H}$ do not commute, we get
\begin{equation}
    \hat{\Delta}_1 = \sum_{k=1}^\infty \frac{(-\beta)^k}{k!} \sum_{i=0}^{k-1} \hat{H}_0^i ~ \Delta\hat{H} ~ \hat{H}_0^{k-(i+1)}
\end{equation}
and
\begin{equation}
    \hat{\Delta}_2 = \sum_{k=2}^\infty \frac{(-\beta)^k}{k!} \sum_{i=0}^{k-2} ~ \sum_{j=0}^{k-(i+2)} \hat{H}_0^i ~ \Delta\hat{H} ~ \hat{H}_0^j ~ \Delta\hat{H} ~ \hat{H}_0^{k-(i+j+2)}
\end{equation}
for the operators $\hat{\Delta}_k$ that are linear ($k=1$) and quadratic ($k=2$) in $\Delta \hat{H}$, respectively.
Combining Eqs. \eqref{eq:deltaF} and \eqref{eq:expansion} leads to the expression
\begin{equation}
    \Delta F = -\frac{1}{\beta} \ln \left( 1 + \frac{\sum_{k=1}^\infty \textrm{tr}[\hat{\Delta}_k]}{\textrm{tr}[\exp(-\beta \hat{H}_0)]} \right) = -\frac{1}{\beta} \ln (1+x).
\label{eq:deltaFApprox}
\end{equation}
It should be clear at this stage that the Taylor expansion defined in Eq. \eqref{eq:expansion} is motivated by approximating $x$, that is,
\begin{equation}
    x = \frac{\sum_{k=1}^\infty \textrm{tr}[\hat{\Delta}_k]}{\textrm{tr}[\exp(-\beta \hat{H}_0)]} \approx 
    	\frac{\textrm{tr}[\hat{\Delta}_1] + \textrm{tr}[\hat{\Delta}_2]}{\textrm{tr}[\exp(-\beta \hat{H}_0)]}
\end{equation}
where terms at most quadratic in the dipole moment $\mu$ are retained (see the Supplementary Material for more information).

In order to approximate $\Delta F$, the traces of $\hat{\Delta}_1$ and $\hat{\Delta}_2$ need to be evaluated, which is carried out
in the basis spanned by the eigenfunctions of $\hat{H}_0$. For the sake of simplicity, we introduce the composite index 
$\alpha = (\alpha_1,\dots,\alpha_N)$ which incorporates every individual molecular eigenstate label. Using this shorthand notation,
molecular eigenstates are readily combined into $|\alpha \rangle = \otimes_{i=1}^N | \alpha_i \rangle$.
Thus, as derived in the Supplementary Material, the trace of $\hat{\Delta}_1$ becomes
\begin{gather}
    \textrm{tr}[\hat{\Delta}_1] = 
    \sum_{\alpha n} \langle \alpha n \big| \sum_{k=1}^\infty \frac{(-\beta)^k}{k!} \sum_{i=0}^{k-1} \hat{H}_0^i ~ \Delta\hat{H} ~ \hat{H}_0^{k-(i+1)} \big| \alpha n \rangle \label{eq:delta1} \\
    = -\frac{\beta g^2 Q_\textrm{c}}{\hbar \omega_\textrm{c}} \sum_{\alpha} \exp (-\beta E_{\alpha}) (\mu^2)_{\alpha\alpha} \nonumber
\end{gather}
where $E_{\alpha} = \sum_{i=1}^N \epsilon_{\alpha_i}$ and the canonical partition function of the cavity mode is equal to
\begin{equation}
    Q_\textrm{c} = \sum_{n=0}^\infty \exp (-\beta \epsilon_n) = \frac{\exp(-\beta \hbar \omega_\textrm{c}/2)}{1-\exp (-\beta \hbar \omega_\textrm{c})}.
\end{equation}
Similarly, one can show (see the Supplementary Material) that
\begin{gather}
    \textrm{tr}[\hat{\Delta}_2] = 
    \sum_{\alpha n} \langle \alpha n \big| \sum_{k=2}^\infty \frac{(-\beta)^k}{k!} \sum_{i=0}^{k-2} ~ \sum_{j=0}^{k-(i+2)} \hat{H}_0^i ~ \Delta\hat{H} ~ \hat{H}_0^j ~ 
        \Delta\hat{H} ~ \hat{H}_0^{k-(i+j+2)} \big| \alpha n \rangle \label{eq:delta2} \\ \nonumber
    = \beta g^2 Q_\textrm{c}^2 \sum_{\alpha\alpha'} \exp(-\beta E_{\alpha}) |\mu_{\alpha\alpha'}|^2
        \left( \frac{\exp(\beta \hbar \omega_\textrm{c}/2)}{E_{\alpha'}-E_{\alpha} + \hbar \omega_\textrm{c}} +
            \frac{\exp(-\beta \hbar \omega_\textrm{c}/2)}{E_{\alpha'}-E_{\alpha} - \hbar \omega_\textrm{c}} \right).
\end{gather}
Derivations presented in the Supplementary Material also reveal that terms linear in $\mu$ drop out from the operator trace $\textrm{tr}[\hat{\Delta}_1]$.

The next step is to reformulate Eqs. \eqref{eq:delta1} and \eqref{eq:delta2} in terms of individual molecular properties (such as energy levels $\epsilon_{\alpha_i}$).
This can be achieved by the decomposition
\begin{equation}
    \mu = \sum_{i=1}^N \mu_i
\label{eq:muDecomp}
\end{equation}
where $\mu$ is expressed as a sum of molecular dipole moments $\mu_i$. In order to simplify further derivations, 
$\textrm{tr}[\hat{\Delta}_1]$ of Eq. \eqref{eq:delta1} is rewritten as
\begin{equation}
    \textrm{tr}[\hat{\Delta}_1] = 
     -\frac{\beta g^2 Q_\textrm{c}}{\hbar \omega_\textrm{c}} \sum_{\alpha \alpha'} \exp (-\beta E_{\alpha}) | \mu_{\alpha\alpha'} |^2
   \label{eq:delta1New}
\end{equation}
where the relation 
$(\mu^2)_{\alpha\alpha} = \langle \alpha | \mu^2 | \alpha \rangle = \sum_{\alpha'} \langle \alpha | \mu | \alpha' \rangle \langle \alpha' | \mu | \alpha \rangle =  
\sum_{\alpha'} | \mu_{\alpha\alpha'} |^2 $ is applied (that is, a resolution of identity is inserted in between the two $\mu$ operators).
With this, $\textrm{tr}[\hat{\Delta}_1]$ and $\textrm{tr}[\hat{\Delta}_2]$ can be combined into
\begin{gather}
    \textrm{tr}[\hat{\Delta}_1] + \textrm{tr}[\hat{\Delta}_2] = 
    \beta g^2 Q_\textrm{c} \sum_{\alpha \alpha'} \exp (-\beta E_{\alpha}) | \mu_{\alpha\alpha'} |^2 \label{eq:delta12} \\ \nonumber
     \left[ -\frac{1}{\hbar \omega_\textrm{c}} + Q_\textrm{c} \left( \frac{\exp(\beta \hbar \omega_\textrm{c}/2)}{E_{\alpha'}-E_{\alpha} + \hbar \omega_\textrm{c}} + 
     \frac{\exp(-\beta \hbar \omega_\textrm{c}/2)}{E_{\alpha'}-E_{\alpha} - \hbar \omega_\textrm{c}} \right) \right].
\end{gather}

According to Eq. \eqref{eq:delta12}, the expression for $\textrm{tr}[\hat{\Delta}_1]+\textrm{tr}[\hat{\Delta}_2]$ involves matrix elements of $\mu$
in the basis of molecular eigenstates. Using Eq. \eqref{eq:muDecomp}, we get
\begin{equation}
    \mu_{\alpha \alpha'} = \sum_{i=1}^N (\mu_i)_{\alpha_i \alpha_i'} \prod_{\substack{j=1 \\ j \neq i}}^N \delta_{\alpha_j \alpha_j'}
   \label{eq:muMx}
\end{equation}
where the Kronecker delta terms appear due to the orthonormality of the molecular eigenbasis. Taking the absolute square of $\mu_{\alpha \alpha'}$ gives
\begin{equation}
    | \mu_{\alpha \alpha'} |^2 = \sum_{i=1}^N \left( (\mu_i)_{\alpha_i \alpha_i'} \right)^2 \prod_{\substack{j=1 \\ j \neq i}}^N \delta_{\alpha_j \alpha_j'} + 
        \sum_{\substack{i,j=1 \\ i \neq j}}^N (\mu_i)_{\alpha_i \alpha_i'} (\mu_j)_{\alpha_j \alpha_j'} \prod_{k=1}^N \delta_{\alpha_k \alpha_k'}
   \label{eq:muMx2}
\end{equation}
where real matrix elements are assumed. Substituting Eq. \eqref{eq:muMx2} into Eq. \eqref{eq:delta12} yields
\begin{gather}
    \textrm{tr}[\hat{\Delta}_1] + \textrm{tr}[\hat{\Delta}_2] = 
    \beta g^2 Q_\textrm{c} \sum_{i=1}^N \sum_{\gamma_i} \exp (-\beta E_{\gamma_i}) \sum_{\alpha_i \alpha_i'} \exp (-\beta \epsilon_{\alpha_i}) 
        ( (\mu_i)_{\alpha_i \alpha_i'} )^2 \label{eq:delta12Detailed} \\ \nonumber
     \left[ -\frac{1}{\hbar \omega_\textrm{c}} + Q_\textrm{c} 
     \left( \frac{\exp(\beta \hbar \omega_\textrm{c}/2)}{\epsilon_{\alpha_i'}-\epsilon_{\alpha_i} + \hbar \omega_\textrm{c}} + 
     \frac{\exp(-\beta \hbar \omega_\textrm{c}/2)}{\epsilon_{\alpha_i'}-\epsilon_{\alpha_i} - \hbar \omega_\textrm{c}} \right) \right] \\ \nonumber
     + \beta g^2 Q_\textrm{c} \sum_{\substack{i,j=1 \\ i \neq j}}^N \sum_\alpha \exp (-\beta E_{\alpha}) (\mu_i)_{\alpha_i \alpha_i} (\mu_j)_{\alpha_j \alpha_j} 
     \frac{Q_\textrm{c} \left[ \exp(\beta \hbar \omega_\textrm{c}/2) - \exp(-\beta \hbar \omega_\textrm{c}/2) \right]-1}{\hbar \omega_\textrm{c}}
\end{gather}
where $\gamma_i = (\alpha_1,\dots,\alpha_{i-1},\alpha_{i+1},\dots,\alpha_N)$ and the relation
$E_{\alpha} = \sum_{k=1}^N \epsilon_{\alpha_k} = E_{\gamma_i} + \epsilon_{\alpha_i}$ is utilised.
It is straightforward to prove the identity
\begin{equation}
    Q_\textrm{c} \left[ \exp(\beta \hbar \omega_\textrm{c}/2) - \exp(-\beta \hbar \omega_\textrm{c}/2) \right] = 1
   \label{eq:QcIdentity}
\end{equation}
which cancels the second term in Eq. \eqref{eq:delta12Detailed}. Thus, Eq. \eqref{eq:delta12Detailed} can be reduced to
\begin{gather}
    \textrm{tr}[\hat{\Delta}_1] + \textrm{tr}[\hat{\Delta}_2] = 
    \beta g^2 N Q_\textrm{c} Q_\textrm{m}^{N-1} \sum_{\alpha_1 \ne \alpha_1'} \exp (-\beta \epsilon_{\alpha_1}) 
        ( (\mu_1)_{\alpha_1 \alpha_1'} )^2 \label{eq:delta12Detailed2} \\ \nonumber
     \left[ -\frac{1}{\hbar \omega_\textrm{c}} + Q_\textrm{c} 
     \left( \frac{\exp(\beta \hbar \omega_\textrm{c}/2)}{\epsilon_{\alpha_1'}-\epsilon_{\alpha_1} + \hbar \omega_\textrm{c}} + 
     \frac{\exp(-\beta \hbar \omega_\textrm{c}/2)}{\epsilon_{\alpha_1'}-\epsilon_{\alpha_1} - \hbar \omega_\textrm{c}} \right) \right]
\end{gather}
where we employ the fact that molecules are identical, $Q_\textrm{m} = \sum_{\alpha_1} \exp (-\beta \epsilon_{\alpha_1})$ is 
the canonical partition function of an isolated molecule and $\sum_{\gamma_1} \exp (-\beta E_{\gamma_1}) = Q_\textrm{m}^{N-1}$.
Note that the summation in Eq. \eqref{eq:delta12Detailed2} can be restricted to $\alpha_1 \ne \alpha_1'$ due to Eq. \eqref{eq:QcIdentity}.

This leads us to the final result
\begin{gather}
    x \approx \frac{\textrm{tr}[\hat{\Delta}_1] + \textrm{tr}[\hat{\Delta}_2]}{Q_0} = \frac{\beta G^2}{Q_\textrm{m}} 
   	 \sum_{\alpha_1 \neq \alpha_1'} \left( (\mu_1)_{\alpha_1 \alpha_1'} \right)^2 \exp (-\beta \epsilon_{\alpha_1}) \label{eq:xFinal} \\ \nonumber
         \left[ -\frac{1}{\hbar \omega_\textrm{c}} + Q_\textrm{c} 
     	\left( \frac{\exp(\beta \hbar \omega_\textrm{c}/2)}{\epsilon_{\alpha_i'}-\epsilon_{\alpha_i} + \hbar \omega_\textrm{c}} + 
     	\frac{\exp(-\beta \hbar \omega_\textrm{c}/2)}{\epsilon_{\alpha_i'}-\epsilon_{\alpha_i} - \hbar \omega_\textrm{c}} \right) \right] 
\end{gather}
where we have defined the collective coupling strength $G = g \sqrt{N}$ and the uncoupled partition function is
$Q_0 = \textrm{tr}[\exp(-\beta \hat{H}_0)] = Q_\textrm{c} Q_\textrm{m}^N$.
We stress that $x$ of Eq. \eqref{eq:xFinal} depends solely on parameters of the cavity mode and properties of a single unperturbed molecule.
Moreover, $x$ is consistent with the approximation which neglects all terms containing $\mu$ exactly $k$ times with $k>2$.
For sufficiently small values of $x$, $\Delta F$ can be approximated by the expansion
\begin{equation}
	\Delta F =  -\frac{1}{\beta} \ln (1+x) \approx -\frac{1}{\beta} \left( x-\frac{x^2}{2} \right)
\end{equation}
which shows that the leading-order term in $\Delta F$ is proportional to $G^2$. Since $Q_0 = Q_\textrm{c} Q_\textrm{m}^N$, we get
\begin{equation}
    F_0 = -\frac{1}{\beta} \ln Q_0 = -\frac{1}{\beta} \left( \ln Q_\textrm{c} + N \ln Q_\textrm{m} \right)
\label{eq:F0gen}
\end{equation}
for the uncoupled cavity-molecule ensemble.

Finally, the problem of potentially resonant terms $1/(\epsilon_{\alpha_1'}-\epsilon_{\alpha_1} \pm \hbar \omega_\textrm{c})$ in Eq. \eqref{eq:xFinal} is addressed. 
For example, for $\epsilon_{\alpha_1} - \epsilon_{\alpha_1'} \approx \hbar \omega_\textrm{c}$, we first realise that $(\mu_1)_{\alpha_1 \alpha_1'} = (\mu_1)_{\alpha_1' \alpha_1}$
(matrix elements are real), then we collect both resonant terms and consider the limit
\begin{gather}
    \lim_{\omega_\textrm{c} \rightarrow (\epsilon_{\alpha_1} - \epsilon_{\alpha_1'})/\hbar} 
        \left( \frac{\exp (-\beta \epsilon_{\alpha_1}) \exp(\beta \hbar \omega_\textrm{c}/2)}{\epsilon_{\alpha_1'} - \epsilon_{\alpha_1} + \hbar \omega_\textrm{c}} +
               \frac{\exp (-\beta \epsilon_{\alpha_1'}) \exp(-\beta \hbar \omega_\textrm{c}/2)}{\epsilon_{\alpha_1} - \epsilon_{\alpha_1'} - \hbar \omega_\textrm{c}}\right) \label{eq:limit} \\ \nonumber
    = \lim_{\omega_\textrm{c} \rightarrow (\epsilon_{\alpha_1} - \epsilon_{\alpha_1'})/\hbar} 
         \frac{\exp (-\beta \epsilon_{\alpha_1}) \exp(\beta \hbar \omega_\textrm{c}/2) - \exp (-\beta \epsilon_{\alpha_1'}) \exp(-\beta \hbar \omega_\textrm{c}/2)}{\epsilon_{\alpha_1'} - \epsilon_{\alpha_1} + \hbar \omega_\textrm{c}} \\ \nonumber
    = \beta \exp \left( -\frac{\beta}{2} (\epsilon_{\alpha_1} + \epsilon_{\alpha_1'}) \right)
\end{gather}
where L'H\^{o}pital's rule is used for an indeterminate form of type $0/0$. Eq. \eqref{eq:limit} clearly demonstrates that a finite limit
is obtained for the case $\omega_\textrm{c} \rightarrow (\epsilon_{\alpha_1} - \epsilon_{\alpha_1'})/\hbar$. The same argument
provides a finite limit for the case $\omega_\textrm{c} \rightarrow (\epsilon_{\alpha_1'} - \epsilon_{\alpha_1})/\hbar$ as well.

\section{Model system}
\label{sec:model}

As a next step, the general theory is applied to a model system which can be solved analytically and thus offers the possibility to
numerically test and validate the formulae outlined in the previous section.
In what follows, we shall determine the canonical partition function $Q$ and Helmholtz free energy correction $\Delta F$ due to cavity-molecule
coupling for an ensemble of $N$ identical one-dimensional harmonic oscillators coupled to a single cavity mode.
To this end, the Hamiltonian of the $i$th molecule assumes the form
\begin{equation}
	\hat{H}^{(0)}_i = \frac{1}{2} \left (\hat{p}_i^2 + \omega^2 q_i^2 \right)
\label{eq:H0_HO}
\end{equation}
where $\omega$ denotes the oscillator frequency and the mass is set to unity. Energy levels of $\hat{H}^{(0)}_i$ are given by
\begin{equation}
	\epsilon_{\alpha_i} = \hbar \omega \left( \alpha_i+\frac{1}{2} \right)
\end{equation}
where the quantum number $\alpha_i = 0,1,2,\dots$ labels energy levels and eigenstates of the one-dimensional harmonic oscillator.
The electric dipole moment of the $i$th molecule is described by a linear function, that is, 
\begin{equation}
	\mu_i = a q_i
\end{equation}
where the slope of $\mu_i$ is determined by the parameter $a$.
Matrix elements of $\mu_i$ in the harmonic oscillator eigenbasis $| \alpha_i \rangle$ can be obtained as
\begin{equation}
	\langle \alpha_i | \mu_i | \alpha'_i \rangle = a \langle \alpha_i | q_i | \alpha'_i \rangle = 
        a \sqrt{\frac{\hbar}{2\omega}} \left( \sqrt{\alpha_i} \delta_{\alpha_i,\alpha'_i+1} +
		 \sqrt{\alpha_i+1} \delta_{\alpha_i,\alpha'_i-1} \right)
\end{equation}
where the standard harmonic oscillator coordinate matrix element formula is used.\cite{77CoDiLa}

It is advantageous to express the total dipole moment
\begin{equation}
    \mu = a \sum_{i=1}^N q_i = a \sqrt{N} Q_1
\end{equation}
as a function of the collective mode $Q_1 = 1/\sqrt{N} \sum_{i=1}^N q_i$ and further introduce $N-1$ orthogonal and degenerate modes $Q_i$ ($i=2,\dots,N$).
It is easy to verify that the coordinate transformation $(q_1,\dots,q_N) \rightarrow (Q_1,\dots,Q_N)$ yields
\begin{gather}
    \hat{H} = \frac{1}{2} \sum_{i=2}^N \left( \hat{P}_i^2 + \omega^2 Q_i^2 \right) + \frac{1}{2} \left( \hat{p}_\textrm{c}^2 + \hat{P}_1^2 \right) +
        \frac{1}{2} \left( \omega_\textrm{c}^2 q_\textrm{c}^2 + \omega^2 Q_1^2 \right) \label{eq:Hdecoupled1} \\ 
    - G a \sqrt{\frac{2 \omega_\textrm{c}}{\hbar}} q_\textrm{c} Q_1 + 
        \frac{G^2 a^2}{\hbar \omega_\textrm{c}} Q_1^2  \nonumber
\end{gather}
for the Hamiltonian of Eq. \eqref{eq:H} and modes $Q_i$ ($i=2,\dots,N$) are essentially decoupled from $q_\textrm{c}$ and $Q_1$.
We recall that $G = g \sqrt{N}$.
In Eq. \eqref{eq:Hdecoupled1} the only remaining coupling between the modes $q_\textrm{c}$ and $Q_1$ is proportional to $q_\textrm{c} Q_1$,
which can be eliminated by diagonalising the two-dimensional force constant matrix
\begin{equation}
    \mathbf{W} =
    \begin{pmatrix}
    \omega_\textrm{c}^2 & ~~ -G a \sqrt{\frac{2 \omega_\textrm{c}}{\hbar}} \\
    -G a \sqrt{\frac{2 \omega_\textrm{c}}{\hbar}} & ~~ \omega^2 + \frac{2 G^2 a^2}{\hbar \omega_\textrm{c}}
    \end{pmatrix}.
\end{equation}
Note that the potential energy containing $q_\textrm{c}$- and $Q_1$-dependent terms in Eq. \eqref{eq:Hdecoupled1} can be
expressed as $V = \frac{1}{2} \mathbf{x}^T \mathbf{W} \mathbf{x}$ with $\mathbf{x}^T = (q_\textrm{c},Q_1)$.
Obviously, the coordinate transformation $(q_\textrm{c}, Q_1) \rightarrow (Q_+,Q_-)$ results in
\begin{equation}
    \hat{H} = \frac{1}{2} \sum_{i=2}^N \left( \hat{P}_i^2 + \omega^2 Q_i^2 \right) + 
        \frac{1}{2} \left( \hat{P}_+^2 + \omega_+^2 Q_+^2 \right) + \frac{1}{2} \left( \hat{P}_-^2 + \omega_-^2 Q_-^2 \right)
\label{eq:Hdecoupled2}
\end{equation}
where the new frequencies $\omega_\pm$ and modes $Q_\pm$ are obtained by diagonalising $\mathbf{W}$.

We are now in a position to evaluate the canonical partition function of the model defined by $\hat{H}$ of Eq. \eqref{eq:Hdecoupled2}.
Since $\hat{H}$ describes $N+1$ uncoupled harmonic oscillator degrees of freedom, we get
\begin{equation}
    Q = Q_{\omega_+} Q_{\omega_-} Q_{\omega}^{N-1} = 
        \frac{\exp(-\beta \hbar \omega_+/2)}{1-\exp (-\beta \hbar \omega_+)}
        \frac{\exp(-\beta \hbar \omega_-/2)}{1-\exp (-\beta \hbar \omega_-)} 
        \left[ \frac{\exp(-\beta \hbar \omega/2)}{1-\exp (-\beta \hbar \omega)} \right]^{N-1}
\end{equation}
for the partition function. If there is no cavity-molecule coupling ($g=0$), the corresponding partition function equals
\begin{equation}
    Q_0 = Q_\textrm{c} Q_{\omega}^N = 
        \frac{\exp(-\beta \hbar \omega_\textrm{c}/2)}{1-\exp (-\beta \hbar \omega_\textrm{c})}
        \left[ \frac{\exp(-\beta \hbar \omega/2)}{1-\exp (-\beta \hbar \omega)} \right]^N.
\end{equation}
With these quantities at hand, the Helmholtz free energy correction $\Delta F$ due to cavity-molecule coupling is readily expressed as
\begin{equation}
    \Delta F = -\frac{1}{\beta} \ln \frac{Q}{Q_0} = -\frac{1}{\beta} \ln \frac{Q_{\omega_+} Q_{\omega_-}}{Q_\textrm{c} Q_{\omega}}
\label{eq:deltaFExact}
\end{equation}
while the free energy of the uncoupled system reads
\begin{equation}
	F_0 = -\frac{1}{\beta} \ln Q_0 = -\frac{1}{\beta} \left( \ln Q_\textrm{c} + N \ln Q_{\omega} \right).
\label{eq:F0}
\end{equation}

\section{Results and Discussion}

In what follows, the performance of the method presented in Section \ref{sec:theory} is evaluated for the model system outlined in Section \ref{sec:model}.
Namely, the approximate Helmholtz free energy correction due to cavity-molecule coupling, $\Delta F_\textrm{approx} = -1/\beta \ln (1+x)$ (see Eq. \eqref{eq:deltaFApprox}
and also Eq. \eqref{eq:xFinal} for the definition of $x$), is compared to its exact counterpart, $\Delta F_\textrm{exact}$ (see Eq. \eqref{eq:deltaFExact}), for various
cavity parameters. Moreover, the ratio $\Delta F_\textrm{exact}/F_0$ (where $F_0$ equals the Helmholtz free energy of the uncoupled cavity-molecule system,
see Eq. \eqref{eq:F0gen} and Eq. \eqref{eq:F0}) is investigated, which sheds light on the magnitude of cavity-induced changes in the thermodynamic properties of the system.
To this end, model parameters $\omega = 1000 ~ \textrm{cm}^{-1}$ and $a = 0.01 ~ \textrm{au}$ are applied throughout this section,
and the cavity wavenumber $\omega_\textrm{c}$ is swept across the interval $[500,1500] ~ \textrm{cm}^{-1}$. 
Since we focus on the collective ultrastrong coupling (cUSC) regime (the collective coupling strength is comparable to $\omega_\textrm{c}$),
we employ $G \le 1000 ~ \textrm{cm}^{-1}$. Finally, we assume that the ensemble comprises a large number of molecules, that is, $N>>1$.

\begin{figure}
\centering
\includegraphics[scale=0.75]{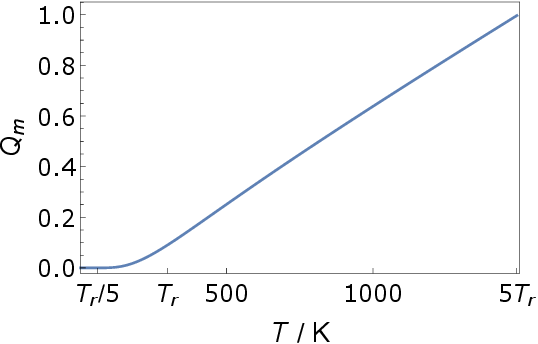}
\includegraphics[scale=0.75]{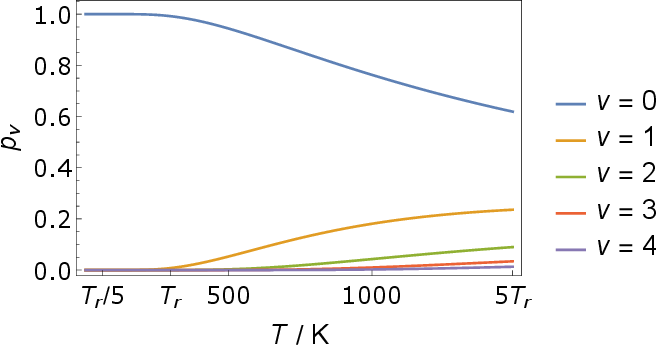}
\caption{Temperature dependence of the molecular partition function $Q_\textrm{m}$ (left panel) and Boltzmann populations
	     $p_v = \exp(-\beta \epsilon_v) / Q_\textrm{m}$ (right panel) with $\beta = 1/(k_\textrm{B}T)$.
	     Populations of the lowest five eigenstates ($v=0,\dots,4$, where $v$ is the vibrational quantum number) are shown
	     for the one-dimensional harmonic oscillator model with $\omega = 1000 ~ \textrm{cm}^{-1}$.
	     $T_\textrm{r} = 298.15 ~ \textrm{K}$ refers to the room temperature.}
\label{fig:qmp}
\end{figure}

Fig. \ref{fig:qmp} shows the temperature dependence of the molecular partition function $Q_\textrm{m}$ and Boltzmann populations $p_v = \exp(-\beta \epsilon_v) / Q_\textrm{m}$
of the lowest five molecular eigenstates ($v = 0,\dots,4$) for the one-dimensional harmonic oscillator model of Section \ref{sec:model}.
As expected, $Q_\textrm{m}$ follows the temperature dependence of the quantum harmonic oscillator partition function and excited eigenstates
get populated as the temperature increases. Guided by the shape of the $Q_\textrm{m}(T)$ and $p_v(T)$ curves, we select
three specific temperatures for further investigation, namely, the room temperature, $T_\textrm{r} = 298.15 ~ \textrm{K}$, $T = T_\textrm{r}/5$ and $T = 5T_\textrm{r}$.

\begin{figure}
\centering
\includegraphics[scale=1.0]{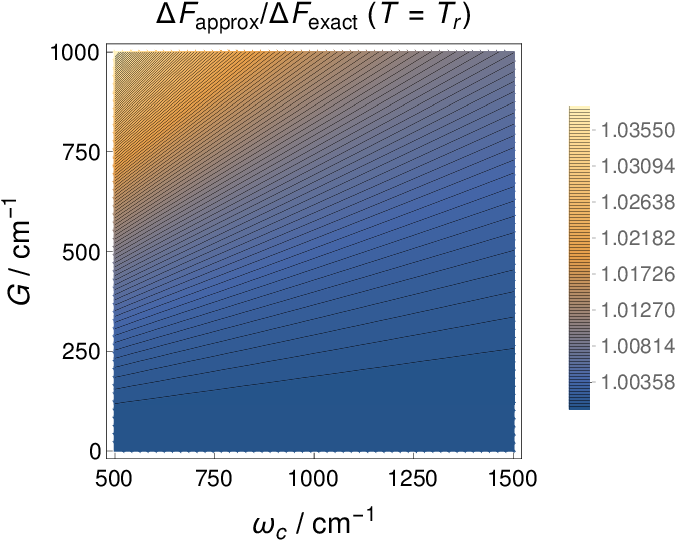}
\caption{Cavity wavenumber ($\omega_\textrm{c}$, in $\textrm{cm}^{-1}$) and collective coupling strength ($G$, in $\textrm{cm}^{-1}$)
	      dependence of the ratio of the approximate and exact Helmholtz free energy corrections 
	      ($\Delta F_\textrm{approx} / \Delta F_\textrm{exact}$) due to cavity-molecule coupling. $T = T_\textrm{r} = 298.15 ~ \textrm{K}$
	      corresponds to the room temperature and molecules are described with the one-dimensional harmonic oscillator model with
	      $\omega = 1000 ~ \textrm{cm}^{-1}$.}
\label{fig:dpdwcg}
\end{figure}

\begin{figure}
\centering
\includegraphics[scale=0.7]{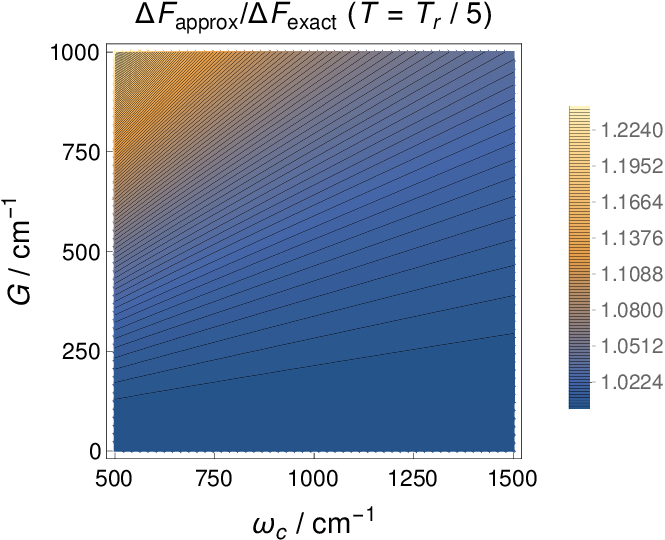}
\includegraphics[scale=0.7]{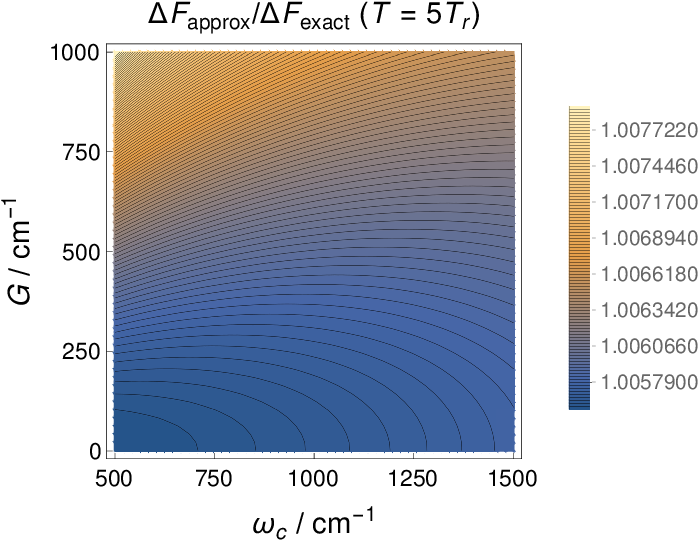}
\caption{Cavity wavenumber ($\omega_\textrm{c}$, in $\textrm{cm}^{-1}$) and collective coupling strength ($G$, in $\textrm{cm}^{-1}$)
              dependence of the ratio of the approximate and exact Helmholtz free energy corrections 
	      ($\Delta F_\textrm{approx} / \Delta F_\textrm{exact}$) due to cavity-molecule coupling. 
	      The left and right panels show results for temperatures $T = T_\textrm{r}/5$ and $T = 5T_\textrm{r}$, respectively,
	      where $T_\textrm{r} = 298.15 ~ \textrm{K}$ corresponds to the room temperature.
	      Molecules are described with the one-dimensional harmonic oscillator model with $\omega = 1000 ~ \textrm{cm}^{-1}$.}
\label{fig:dpdwcg2}
\end{figure}

\begin{figure}
\centering
\includegraphics[scale=1.0]{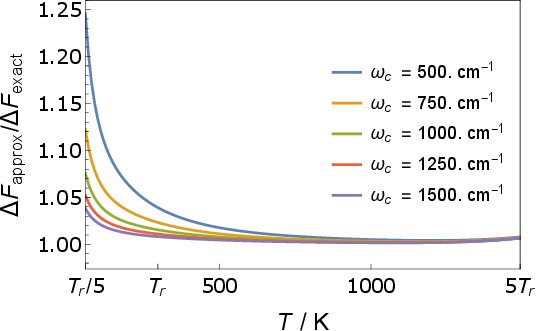}
\caption{Temperature dependence of the ratio of the approximate and exact Helmholtz free energy corrections 
	     ($\Delta F_\textrm{approx} / \Delta F_\textrm{exact}$) due to cavity-molecule coupling. $T_\textrm{r} = 298.15 ~ \textrm{K}$
	     corresponds to the room temperature, molecules are described with the one-dimensional harmonic oscillator model with
	     $\omega = 1000 ~ \textrm{cm}^{-1}$ and the collective coupling strength equals $G = 1000 ~ \textrm{cm}^{-1}$.
	     Results for different cavity wavenumbers $\omega_\textrm{c}$ are shown in different colours as indicated by the plot legend.}
\label{fig:dpdt}
\end{figure}

Fig. \ref{fig:dpdwcg} depicts the dependence of $\Delta F_\textrm{approx} / \Delta F_\textrm{exact}$ on the cavity parameters $\omega_\textrm{c}$ and $G$ for $T = T_\textrm{r}$.
As seen in Fig. \ref{fig:dpdwcg}, $\Delta F_\textrm{approx}$ systematically overestimates $\Delta F_\textrm{exact}$ with an error less or equal than a few percent.
In addition, the error of approximating $\Delta F_\textrm{exact}$ increases as $\omega_\textrm{c}$ decreases and $G$ increases.
Similar observations can be made for $T = T_\textrm{r}/5$ and $T = 5T_\textrm{r}$, see Fig. \ref{fig:dpdwcg2}, which also gives an idea about the performance of
the approximation at different temperatures.
Comparison of the results displayed in Fig. \ref{fig:dpdwcg} and Fig. \ref{fig:dpdwcg2} reveals that the error of the approximation is higher at lower temperatures.
One can also notice in the right panel of Fig. \ref{fig:dpdwcg2} that $\Delta F_\textrm{approx} / \Delta F_\textrm{exact}$
for $T = 5T_\textrm{r}$ exhibits a more complex contour structure than for $T = T_\textrm{r}$ and $T = T_\textrm{r}/5$, which can be attributed
to appreciable excited eigenstate populations at $T = 5T_\textrm{r}$ (see the right panel of Fig. \ref{fig:qmp}).
Similar conclusions can be drawn by inspecting Fig. \ref{fig:dpdt} where $\Delta F_\textrm{approx} / \Delta F_\textrm{exact}$ is shown as a function of temperature
for five different values of $\omega_\textrm{c}$ and $G = 1000 ~ \textrm{cm}^{-1}$. Again, it is conspicuous in Fig. \ref{fig:dpdt} that the performance of the
approximation deteriorates with decreasing temperature, reaching about $20 \%$ for $\omega_\textrm{c} = 500 ~ \textrm{cm}^{-1}$ at $T = T_\textrm{r}/5$.

\begin{figure}
\centering
\includegraphics[scale=1.0]{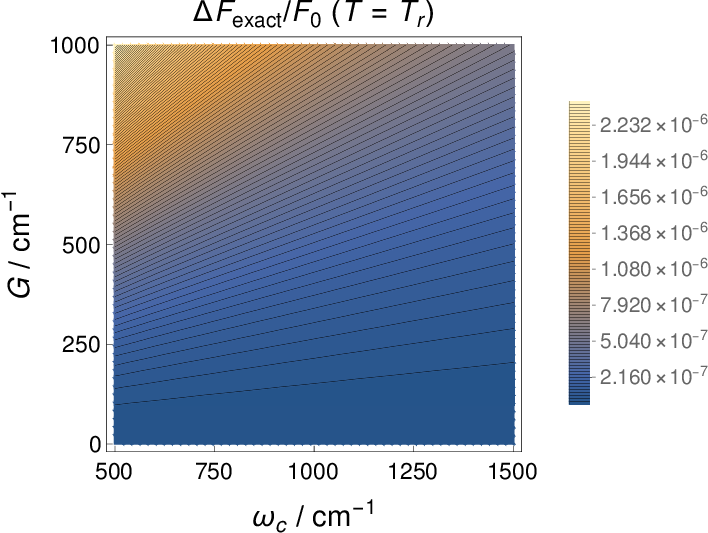}
\caption{Cavity wavenumber ($\omega_\textrm{c}$, in $\textrm{cm}^{-1}$) and collective coupling strength ($G$, in $\textrm{cm}^{-1}$)
              dependence of the ratio of the exact Helmholtz free energy correction due to cavity-molecule coupling and Helmholtz free energy of the uncoupled 
	      cavity-molecule system ($\Delta F_\textrm{exact} / F_0$). $T = T_\textrm{r} = 298.15 ~ \textrm{K}$
	      corresponds to the room temperature, the number of molecules is set to $N=10000$ and
	      molecules are described with the one-dimensional harmonic oscillator model with $\omega = 1000 ~ \textrm{cm}^{-1}$.}
\label{fig:dpf0wcg}
\end{figure}

\begin{figure}
\centering
\includegraphics[scale=0.65]{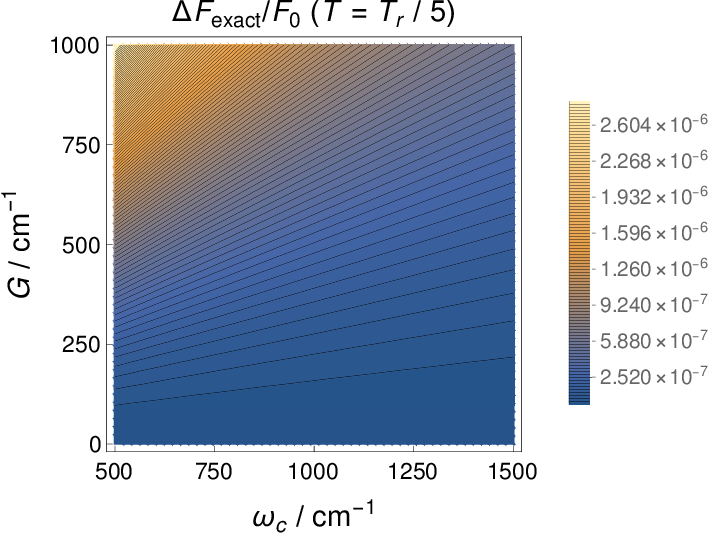}
\includegraphics[scale=0.65]{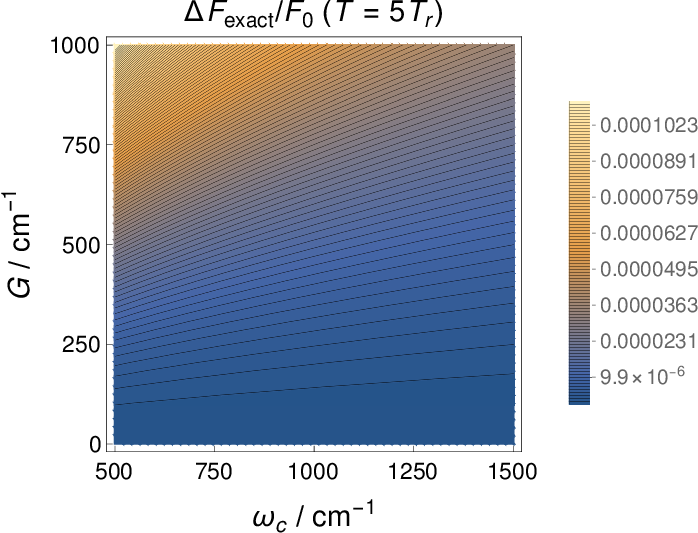}
\caption{Cavity wavenumber ($\omega_\textrm{c}$, in $\textrm{cm}^{-1}$) and collective coupling strength ($G$, in $\textrm{cm}^{-1}$)
              dependence of the ratio of the exact Helmholtz free energy correction due to cavity-molecule coupling and Helmholtz free energy of the uncoupled 
	      cavity-molecule system ($\Delta F_\textrm{exact} / F_0$). 
	      The left and right panels show results for temperatures $T = T_\textrm{r}/5$ and $T = 5T_\textrm{r}$, respectively,
	      where $T_\textrm{r} = 298.15 ~ \textrm{K}$ corresponds to the room temperature.
	      The number of molecules is set to $N=10000$ and
	      molecules are described with the one-dimensional harmonic oscillator model with $\omega = 1000 ~ \textrm{cm}^{-1}$.}
\label{fig:dpf0wcg2}
\end{figure}

\begin{figure}
\centering
\includegraphics[scale=1.0]{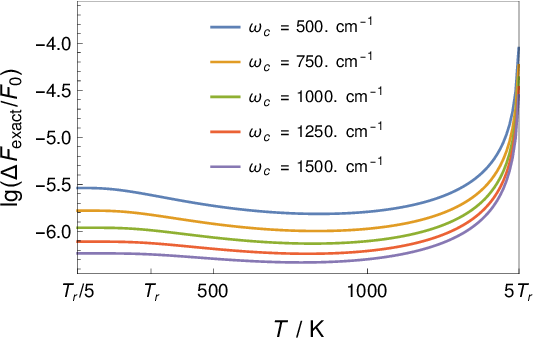}
\caption{Temperature dependence of the ratio of the exact Helmholtz free energy correction due to cavity-molecule coupling and Helmholtz free energy
              of the uncoupled cavity-molecule system ($\Delta F_\textrm{exact} / F_0$, base-10 logarithm is displayed).
              $T_\textrm{r} = 298.15 ~ \textrm{K}$ corresponds to the room temperature and the collective coupling strength equals $G = 1000 ~ \textrm{cm}^{-1}$.
	     The number of molecules is set to $N=10000$ and molecules are described with the one-dimensional harmonic oscillator model with $\omega = 1000 ~ \textrm{cm}^{-1}$.
	     Results for different cavity wavenumbers $\omega_\textrm{c}$ are shown in different colours as indicated by the plot legend.}
\label{fig:dpf0t}
\end{figure}

Fig. \ref{fig:dpf0wcg} presents the dependence of $\Delta F_\textrm{exact} / F_0$ on the cavity parameters $\omega_\textrm{c}$ and $G$ for 
$N = 10000$ molecules at $T = T_\textrm{r}$. One can conclude that lower $\omega_\textrm{c}$ and higher $G$ values both entail higher $\Delta F_\textrm{exact} / F_0$
corrections. In the $\omega_\textrm{c}$ and $G$ ranges shown in Fig. \ref{fig:dpf0wcg} $\Delta F_\textrm{exact} / F_0$ remains on the order
of $10^{-7} - 10^{-6}$, clearly indicating that collective coupling to the cavity mode gives rise to minor changes in the Helmholtz free energy
in this particular case. Again, Fig. \ref{fig:dpf0wcg2} shows results similar to the ones in Fig. \ref{fig:dpf0wcg} for $T = T_\textrm{r}/5$ and $T = 5T_\textrm{r}$. 
Fig. \ref{fig:dpf0t} displays $\Delta F_\textrm{exact} / F_0$ on base-10 logarithmic scale as a function of temperature for five
different values of $\omega_\textrm{c}$ with $N=10000$ and $G = 1000 ~ \textrm{cm}^{-1}$. All curves in Fig. \ref{fig:dpf0t}
have a similar shape, showing minima around the midpoint of the temperature range $T_\textrm{r}/5 < T < 5T_\textrm{r}$.
In addition, it is also visible in Fig. \ref{fig:dpf0t} that $\Delta F_\textrm{exact} / F_0$ can reach $10^{-4}$ at $T = 5T_\textrm{r}$.

\begin{figure}
\centering
\includegraphics[scale=1.0]{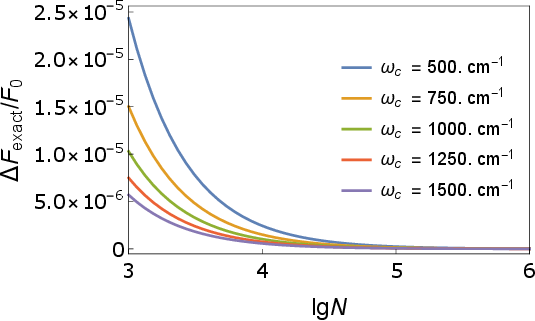}
\caption{Dependence of the ratio of the exact Helmholtz free energy correction due to cavity-molecule coupling and Helmholtz free energy of the uncoupled 
	     cavity-molecule system ($\Delta F_\textrm{exact} / F_0$) on the number of molecules $N$ ($\textrm{lg}N$ denotes the base-10 logarithm of $N$).
	     The temperature is set to $T = T_\textrm{r} = 298.15 ~ \textrm{K}$ (room temperature), molecules are described with the one-dimensional harmonic oscillator model
	     with $\omega = 1000 ~ \textrm{cm}^{-1}$ and the collective coupling strength is kept fixed at $G = 1000 ~ \textrm{cm}^{-1}$.
	     Results for different cavity wavenumbers $\omega_\textrm{c}$ are shown in different colours as indicated by the plot legend.}
\label{fig:dpf0n}
\end{figure}

Last, Fig. \ref{fig:dpf0n} highlights the relationship between $\Delta F_\textrm{exact} / F_0$ and $N$ for five different $\omega_\textrm{c}$ values at $T = T_\textrm{r}$.
Note that the collective coupling strength is kept fixed at a value of $G = 1000 ~ \textrm{cm}^{-1}$ when increasing $N$.
Fig. \ref{fig:dpf0n} shows that one can expect higher $\Delta F_\textrm{exact} / F_0$ corrections for lower values of $\omega_\textrm{c}$.
Furthermore, $\Delta F_\textrm{exact} / F_0$ tends to zero as the number of molecules increases, which can be rationalised as follows.
Both $\Delta F_\textrm{exact}$ and $\Delta F_\textrm{approx}$ depend on $N$ through $G$. Since $G$ is kept fixed in Fig. \ref{fig:dpf0n}, $\Delta F_\textrm{exact}$
remains constant if $N$ changes. On the contrary, $F_0$ scales linearly with $N$ (see Eq. \eqref{eq:F0gen}), which implies that
$\Delta F_\textrm{exact} / F_0 \rightarrow 0$ if $N \rightarrow \infty$ with constant $G$.

Finally, our findings are summarised for the cUSC regime, also in the context of the conclusions drawn in Ref. \onlinecite{20PiBeRa}.
In our results no resonance effects are visible for $\omega_\textrm{c} \approx \omega$. The leading-order correction to $F$ due to cavity-molecule coupling is
proportional to the square of the collective coupling strength, $\Delta F \sim G^2$. Moreover, $\Delta F$ turns out to be positive and $\Delta F$
depends on $N$ (number of molecules) through $G$. Thus, the Helmholtz free energy correction per molecule tends to zero if the number of molecules
goes to infinity while $G$ is kept fixed, that is, $\Delta F / N \rightarrow 0$ for $N \rightarrow \infty$.
For large values of $N$, individual cavity-molecule couplings are weak and the cavity mode has a minor effect on the thermodynamic properties of the ensemble.
This finding is supported by Section \ref{sec:model} where it has been shown that the cavity mode is coupled to a single collective molecular mode $Q_1$ while
collective modes orthogonal to $Q_1$ do not interact with the cavity mode.
All of these observations are in line with the ones reported in Ref. \onlinecite{20PiBeRa} for an ensemble of two-level dipoles coupled to a cavity mode.

\section{Summary and Conclusions}

We have presented a method aimed at the statistical mechanics of molecules coupled to a quantised mode of an infrared cavity.
Namely, a canonical ensemble is assumed and approximate formulae are devised for the canonical partition function $Q$ and Helmholtz free energy $F$ of the ensemble,
which enables the derivation of experimentally measurable thermodynamic quantities. Collective effects arising from cavity-molecule interactions are taken into
account in a rigorous way. We also stress that the method presented is able to treat an arbitrarily large number of molecules.

The core idea of our approach is summarised as follows. First, the Hamiltonian of the ensemble, including the cavity mode, is partitioned.
The first term of this partition is the sum of unperturbed molecular Hamiltonians and the Hamiltonian of the cavity mode, while cavity-molecule interactions
are transformed into the second term. Note that the partition of the Hamiltonian involves no approximations.
Then, the canonical density operator is expanded in terms of the overall dipole moment $\mu$ up to second order, which enables the evaluation of
$Q$ and $F$. It is also shown that expansion terms linear in $\mu$ do not contribute to $Q$ and $F$, and the leading-order correction to $F$ is
proportional to the square of the collective coupling strength.

The performance of the method is tested for a model system of $N$ identical one-dimensional harmonic oscillators coupled to a cavity mode, amenable to an analytic solution.
Numerical tests employing our method have been carried out with various cavity parameters and multiple temperatures for large values of $N$.
Comparison of these results to their exact counterparts demonstrate that our method provides an accurate description of cavity-induced thermodynamic effects in
the collective ultrastrong coupling (cUSC) regime (that is, the collective coupling strength is comparable to the frequency of the cavity mode). 
It is shown that the cavity mode does not have a significant impact on the thermodynamic properties of the ensemble in the cUSC regime,
which is in line with earlier results reported in Ref. \onlinecite{20PiBeRa}.
Moreover, no obvious resonance effects can be observed in our results if the cavity frequency is tuned into resonance with the eigenfrequency of the molecular oscillators.

Although the present work investigates a simple model system, it is obviously possible to employ the current approach with more sophisticated molecular (ro)vibrational models,
ranging from multidimensional harmonic-oscillator and rigid-rotor approximations to highly-accurate variational techniques.
This claim is based on the fact that corrections to $Q$ and $F$ are determined by cavity parameters and properties of an isolated molecule.
The latter can be readily obtained with remarkable accuracy in the fourth age of quantum chemistry.\cite{12CsFaSz}

\bibliographystyle{apsrev4-2}
\bibliography{polstatmech}

\begin{thebibliography}{101}%
\makeatletter
\providecommand \@ifxundefined [1]{%
 \@ifx{#1\undefined}
}%
\providecommand \@ifnum [1]{%
 \ifnum #1\expandafter \@firstoftwo
 \else \expandafter \@secondoftwo
 \fi
}%
\providecommand \@ifx [1]{%
 \ifx #1\expandafter \@firstoftwo
 \else \expandafter \@secondoftwo
 \fi
}%
\providecommand \natexlab [1]{#1}%
\providecommand \enquote  [1]{``#1''}%
\providecommand \bibnamefont  [1]{#1}%
\providecommand \bibfnamefont [1]{#1}%
\providecommand \citenamefont [1]{#1}%
\providecommand \href@noop [0]{\@secondoftwo}%
\providecommand \href [0]{\begingroup \@sanitize@url \@href}%
\providecommand \@href[1]{\@@startlink{#1}\@@href}%
\providecommand \@@href[1]{\endgroup#1\@@endlink}%
\providecommand \@sanitize@url [0]{\catcode `\\12\catcode `\$12\catcode
  `\&12\catcode `\#12\catcode `\^12\catcode `\_12\catcode `\%12\relax}%
\providecommand \@@startlink[1]{}%
\providecommand \@@endlink[0]{}%
\providecommand \url  [0]{\begingroup\@sanitize@url \@url }%
\providecommand \@url [1]{\endgroup\@href {#1}{\urlprefix }}%
\providecommand \urlprefix  [0]{URL }%
\providecommand \Eprint [0]{\href }%
\providecommand \doibase [0]{https://doi.org/}%
\providecommand \selectlanguage [0]{\@gobble}%
\providecommand \bibinfo  [0]{\@secondoftwo}%
\providecommand \bibfield  [0]{\@secondoftwo}%
\providecommand \translation [1]{[#1]}%
\providecommand \BibitemOpen [0]{}%
\providecommand \bibitemStop [0]{}%
\providecommand \bibitemNoStop [0]{.\EOS\space}%
\providecommand \EOS [0]{\spacefactor3000\relax}%
\providecommand \BibitemShut  [1]{\csname bibitem#1\endcsname}%
\let\auto@bib@innerbib\@empty
%</preamble>
\bibitem [{\citenamefont {Hutchison}\ \emph {et~al.}(2012)\citenamefont
  {Hutchison}, \citenamefont {Schwartz}, \citenamefont {Genet}, \citenamefont
  {Devaux},\ and\ \citenamefont {Ebbesen}}]{12HuScGe}%
  \BibitemOpen
  \bibfield  {author} {\bibinfo {author} {\bibfnamefont {J.~A.}\ \bibnamefont
  {Hutchison}}, \bibinfo {author} {\bibfnamefont {T.}~\bibnamefont {Schwartz}},
  \bibinfo {author} {\bibfnamefont {C.}~\bibnamefont {Genet}}, \bibinfo
  {author} {\bibfnamefont {E.}~\bibnamefont {Devaux}},\ and\ \bibinfo {author}
  {\bibfnamefont {T.~W.}\ \bibnamefont {Ebbesen}},\ }\href
  {https://doi.org/10.1002/anie.201107033} {\bibfield  {journal} {\bibinfo
  {journal} {Angew. Chem. Int. Ed.}\ }\textbf {\bibinfo {volume} {51}},\
  \bibinfo {pages} {1592} (\bibinfo {year} {2012})}\BibitemShut {NoStop}%
\bibitem [{\citenamefont {George}\ \emph {et~al.}(2015)\citenamefont {George},
  \citenamefont {Shalabney}, \citenamefont {Hutchison}, \citenamefont {Genet},\
  and\ \citenamefont {Ebbesen}}]{15GeShHu}%
  \BibitemOpen
  \bibfield  {author} {\bibinfo {author} {\bibfnamefont {J.}~\bibnamefont
  {George}}, \bibinfo {author} {\bibfnamefont {A.}~\bibnamefont {Shalabney}},
  \bibinfo {author} {\bibfnamefont {J.~A.}\ \bibnamefont {Hutchison}}, \bibinfo
  {author} {\bibfnamefont {C.}~\bibnamefont {Genet}},\ and\ \bibinfo {author}
  {\bibfnamefont {T.~W.}\ \bibnamefont {Ebbesen}},\ }\href
  {https://doi.org/10.1021/acs.jpclett.5b00204} {\bibfield  {journal} {\bibinfo
   {journal} {J. Phys. Chem. Lett.}\ }\textbf {\bibinfo {volume} {6}},\
  \bibinfo {pages} {1027} (\bibinfo {year} {2015})}\BibitemShut {NoStop}%
\bibitem [{\citenamefont {Shalabney}\ \emph {et~al.}(2015)\citenamefont
  {Shalabney}, \citenamefont {George}, \citenamefont {Hutchison}, \citenamefont
  {Pupillo}, \citenamefont {Genet},\ and\ \citenamefont {Ebbesen}}]{15ShGeHu}%
  \BibitemOpen
  \bibfield  {author} {\bibinfo {author} {\bibfnamefont {A.}~\bibnamefont
  {Shalabney}}, \bibinfo {author} {\bibfnamefont {J.}~\bibnamefont {George}},
  \bibinfo {author} {\bibfnamefont {J.}~\bibnamefont {Hutchison}}, \bibinfo
  {author} {\bibfnamefont {G.}~\bibnamefont {Pupillo}}, \bibinfo {author}
  {\bibfnamefont {C.}~\bibnamefont {Genet}},\ and\ \bibinfo {author}
  {\bibfnamefont {T.~W.}\ \bibnamefont {Ebbesen}},\ }\href
  {https://doi.org/10.1038/ncomms6981} {\bibfield  {journal} {\bibinfo
  {journal} {Nat. Commun.}\ }\textbf {\bibinfo {volume} {6}},\ \bibinfo {pages}
  {5981} (\bibinfo {year} {2015})}\BibitemShut {NoStop}%
\bibitem [{\citenamefont {Ebbesen}(2016)}]{16Ebbesen}%
  \BibitemOpen
  \bibfield  {author} {\bibinfo {author} {\bibfnamefont {T.}~\bibnamefont
  {Ebbesen}},\ }\href {https://doi.org/10.1021/acs.accounts.6b00295} {\bibfield
   {journal} {\bibinfo  {journal} {Acc. Chem. Res.}\ }\textbf {\bibinfo
  {volume} {49}},\ \bibinfo {pages} {2403} (\bibinfo {year}
  {2016})}\BibitemShut {NoStop}%
\bibitem [{\citenamefont {Chikkaraddy}\ \emph {et~al.}(2016)\citenamefont
  {Chikkaraddy}, \citenamefont {De~Nijs}, \citenamefont {Benz}, \citenamefont
  {Barrow}, \citenamefont {Scherman}, \citenamefont {Rosta}, \citenamefont
  {Demetriadou}, \citenamefont {Fox}, \citenamefont {Hess},\ and\ \citenamefont
  {Baumberg}}]{16ChNiBe}%
  \BibitemOpen
  \bibfield  {author} {\bibinfo {author} {\bibfnamefont {R.}~\bibnamefont
  {Chikkaraddy}}, \bibinfo {author} {\bibfnamefont {B.}~\bibnamefont
  {De~Nijs}}, \bibinfo {author} {\bibfnamefont {F.}~\bibnamefont {Benz}},
  \bibinfo {author} {\bibfnamefont {S.}~\bibnamefont {Barrow}}, \bibinfo
  {author} {\bibfnamefont {O.}~\bibnamefont {Scherman}}, \bibinfo {author}
  {\bibfnamefont {E.}~\bibnamefont {Rosta}}, \bibinfo {author} {\bibfnamefont
  {A.}~\bibnamefont {Demetriadou}}, \bibinfo {author} {\bibfnamefont
  {P.}~\bibnamefont {Fox}}, \bibinfo {author} {\bibfnamefont {O.}~\bibnamefont
  {Hess}},\ and\ \bibinfo {author} {\bibfnamefont {J.}~\bibnamefont
  {Baumberg}},\ }\href {https://doi.org/10.1038/nature17974} {\bibfield
  {journal} {\bibinfo  {journal} {Nature}\ }\textbf {\bibinfo {volume} {535}},\
  \bibinfo {pages} {127} (\bibinfo {year} {2016})}\BibitemShut {NoStop}%
\bibitem [{\citenamefont {Schachenmayer}\ \emph {et~al.}(2015)\citenamefont
  {Schachenmayer}, \citenamefont {Genes}, \citenamefont {Tignone},\ and\
  \citenamefont {Pupillo}}]{15ScGeTi}%
  \BibitemOpen
  \bibfield  {author} {\bibinfo {author} {\bibfnamefont {J.}~\bibnamefont
  {Schachenmayer}}, \bibinfo {author} {\bibfnamefont {C.}~\bibnamefont
  {Genes}}, \bibinfo {author} {\bibfnamefont {E.}~\bibnamefont {Tignone}},\
  and\ \bibinfo {author} {\bibfnamefont {G.}~\bibnamefont {Pupillo}},\ }\href
  {https://doi.org/10.1103/PhysRevLett.114.196403} {\bibfield  {journal}
  {\bibinfo  {journal} {Phys. Rev. Lett.}\ }\textbf {\bibinfo {volume} {114}},\
  \bibinfo {pages} {196403} (\bibinfo {year} {2015})}\BibitemShut {NoStop}%
\bibitem [{\citenamefont {Kowalewski}\ \emph {et~al.}(2016)\citenamefont
  {Kowalewski}, \citenamefont {Bennett},\ and\ \citenamefont
  {Mukamel}}]{16KoBeMu}%
  \BibitemOpen
  \bibfield  {author} {\bibinfo {author} {\bibfnamefont {M.}~\bibnamefont
  {Kowalewski}}, \bibinfo {author} {\bibfnamefont {K.}~\bibnamefont
  {Bennett}},\ and\ \bibinfo {author} {\bibfnamefont {S.}~\bibnamefont
  {Mukamel}},\ }\href {https://doi.org/10.1021/acs.jpclett.6b00864} {\bibfield
  {journal} {\bibinfo  {journal} {J. Phys. Chem. Lett.}\ }\textbf {\bibinfo
  {volume} {7}},\ \bibinfo {pages} {2050} (\bibinfo {year} {2016})}\BibitemShut
  {NoStop}%
\bibitem [{\citenamefont {Luk}\ \emph {et~al.}(2017)\citenamefont {Luk},
  \citenamefont {Feist}, \citenamefont {Toppari},\ and\ \citenamefont
  {Groenhof}}]{17LuFeTo}%
  \BibitemOpen
  \bibfield  {author} {\bibinfo {author} {\bibfnamefont {H.~L.}\ \bibnamefont
  {Luk}}, \bibinfo {author} {\bibfnamefont {J.}~\bibnamefont {Feist}}, \bibinfo
  {author} {\bibfnamefont {J.~J.}\ \bibnamefont {Toppari}},\ and\ \bibinfo
  {author} {\bibfnamefont {G.}~\bibnamefont {Groenhof}},\ }\href@noop {}
  {\bibfield  {journal} {\bibinfo  {journal} {J. Chem. Theory Comput.}\
  }\textbf {\bibinfo {volume} {13}},\ \bibinfo {pages} {4324} (\bibinfo {year}
  {2017})}\BibitemShut {NoStop}%
\bibitem [{\citenamefont {Hagenm\"uller}\ \emph {et~al.}(2017)\citenamefont
  {Hagenm\"uller}, \citenamefont {Schachenmayer}, \citenamefont {Sch\"utz},
  \citenamefont {Genes},\ and\ \citenamefont {Pupillo}}]{17HaScSc}%
  \BibitemOpen
  \bibfield  {author} {\bibinfo {author} {\bibfnamefont {D.}~\bibnamefont
  {Hagenm\"uller}}, \bibinfo {author} {\bibfnamefont {J.}~\bibnamefont
  {Schachenmayer}}, \bibinfo {author} {\bibfnamefont {S.}~\bibnamefont
  {Sch\"utz}}, \bibinfo {author} {\bibfnamefont {C.}~\bibnamefont {Genes}},\
  and\ \bibinfo {author} {\bibfnamefont {G.}~\bibnamefont {Pupillo}},\ }\href
  {https://doi.org/10.1103/PhysRevLett.119.223601} {\bibfield  {journal}
  {\bibinfo  {journal} {Phys. Rev. Lett.}\ }\textbf {\bibinfo {volume} {119}},\
  \bibinfo {pages} {223601} (\bibinfo {year} {2017})}\BibitemShut {NoStop}%
\bibitem [{\citenamefont {Hagenm\"uller}\ \emph {et~al.}(2018)\citenamefont
  {Hagenm\"uller}, \citenamefont {Sch\"utz}, \citenamefont {Schachenmayer},
  \citenamefont {Genes},\ and\ \citenamefont {Pupillo}}]{18HaScSc}%
  \BibitemOpen
  \bibfield  {author} {\bibinfo {author} {\bibfnamefont {D.}~\bibnamefont
  {Hagenm\"uller}}, \bibinfo {author} {\bibfnamefont {S.}~\bibnamefont
  {Sch\"utz}}, \bibinfo {author} {\bibfnamefont {J.}~\bibnamefont
  {Schachenmayer}}, \bibinfo {author} {\bibfnamefont {C.}~\bibnamefont
  {Genes}},\ and\ \bibinfo {author} {\bibfnamefont {G.}~\bibnamefont
  {Pupillo}},\ }\href {https://doi.org/10.1103/PhysRevB.97.205303} {\bibfield
  {journal} {\bibinfo  {journal} {Phys. Rev. B}\ }\textbf {\bibinfo {volume}
  {97}},\ \bibinfo {pages} {205303} (\bibinfo {year} {2018})}\BibitemShut
  {NoStop}%
\bibitem [{\citenamefont {Fregoni}\ \emph {et~al.}(2018)\citenamefont
  {Fregoni}, \citenamefont {Granucci}, \citenamefont {Coccia}, \citenamefont
  {Persico},\ and\ \citenamefont {Corni}}]{18FrGrCo}%
  \BibitemOpen
  \bibfield  {author} {\bibinfo {author} {\bibfnamefont {J.}~\bibnamefont
  {Fregoni}}, \bibinfo {author} {\bibfnamefont {G.}~\bibnamefont {Granucci}},
  \bibinfo {author} {\bibfnamefont {E.}~\bibnamefont {Coccia}}, \bibinfo
  {author} {\bibfnamefont {M.}~\bibnamefont {Persico}},\ and\ \bibinfo {author}
  {\bibfnamefont {S.}~\bibnamefont {Corni}},\ }\href
  {https://doi.org/10.1038/s41467-018-06971-y} {\bibfield  {journal} {\bibinfo
  {journal} {Nat. Commun.}\ }\textbf {\bibinfo {volume} {9}},\ \bibinfo {pages}
  {4688} (\bibinfo {year} {2018})}\BibitemShut {NoStop}%
\bibitem [{\citenamefont {Mart{\'i}nez-Mart{\'i}nez}\ \emph
  {et~al.}(2018)\citenamefont {Mart{\'i}nez-Mart{\'i}nez}, \citenamefont {Du},
  \citenamefont {Ribeiro}, \citenamefont {K\'ena-Cohen},\ and\ \citenamefont
  {Yuen-Zhou}}]{18MaDuRi}%
  \BibitemOpen
  \bibfield  {author} {\bibinfo {author} {\bibfnamefont {L.~A.}\ \bibnamefont
  {Mart{\'i}nez-Mart{\'i}nez}}, \bibinfo {author} {\bibfnamefont
  {M.}~\bibnamefont {Du}}, \bibinfo {author} {\bibfnamefont {R.~F.}\
  \bibnamefont {Ribeiro}}, \bibinfo {author} {\bibfnamefont {S.}~\bibnamefont
  {K\'ena-Cohen}},\ and\ \bibinfo {author} {\bibfnamefont {J.}~\bibnamefont
  {Yuen-Zhou}},\ }\href {https://doi.org/10.1021/acs.jpclett.8b00008}
  {\bibfield  {journal} {\bibinfo  {journal} {J. Phys. Chem. Lett.}\ }\textbf
  {\bibinfo {volume} {9}},\ \bibinfo {pages} {1951} (\bibinfo {year}
  {2018})}\BibitemShut {NoStop}%
\bibitem [{\citenamefont {Du}\ \emph {et~al.}(2018)\citenamefont {Du},
  \citenamefont {Mart{\'i}nez-Mart{\'i}nez}, \citenamefont {Ribeiro},
  \citenamefont {Hu}, \citenamefont {Menon},\ and\ \citenamefont
  {Yuen-Zhou}}]{18DuMaRi}%
  \BibitemOpen
  \bibfield  {author} {\bibinfo {author} {\bibfnamefont {M.}~\bibnamefont
  {Du}}, \bibinfo {author} {\bibfnamefont {L.~A.}\ \bibnamefont
  {Mart{\'i}nez-Mart{\'i}nez}}, \bibinfo {author} {\bibfnamefont {R.~F.}\
  \bibnamefont {Ribeiro}}, \bibinfo {author} {\bibfnamefont {Z.}~\bibnamefont
  {Hu}}, \bibinfo {author} {\bibfnamefont {V.~M.}\ \bibnamefont {Menon}},\ and\
  \bibinfo {author} {\bibfnamefont {J.}~\bibnamefont {Yuen-Zhou}},\ }\href
  {https://doi.org/10.1039/C8SC00171E} {\bibfield  {journal} {\bibinfo
  {journal} {Chem. Sci.}\ }\textbf {\bibinfo {volume} {9}},\ \bibinfo {pages}
  {6659} (\bibinfo {year} {2018})}\BibitemShut {NoStop}%
\bibitem [{\citenamefont {Vendrell}(2018)}]{18Vendrell}%
  \BibitemOpen
  \bibfield  {author} {\bibinfo {author} {\bibfnamefont {O.}~\bibnamefont
  {Vendrell}},\ }\href {https://doi.org/10.1103/PhysRevLett.121.253001}
  {\bibfield  {journal} {\bibinfo  {journal} {Phys. Rev. Lett.}\ }\textbf
  {\bibinfo {volume} {121}},\ \bibinfo {pages} {253001} (\bibinfo {year}
  {2018})}\BibitemShut {NoStop}%
\bibitem [{\citenamefont {Rokaj}\ \emph {et~al.}(2018)\citenamefont {Rokaj},
  \citenamefont {Welakuh}, \citenamefont {Ruggenthaler},\ and\ \citenamefont
  {Rubio}}]{18RoWeRu}%
  \BibitemOpen
  \bibfield  {author} {\bibinfo {author} {\bibfnamefont {V.}~\bibnamefont
  {Rokaj}}, \bibinfo {author} {\bibfnamefont {D.~M.}\ \bibnamefont {Welakuh}},
  \bibinfo {author} {\bibfnamefont {M.}~\bibnamefont {Ruggenthaler}},\ and\
  \bibinfo {author} {\bibfnamefont {A.}~\bibnamefont {Rubio}},\ }\href
  {https://doi.org/10.1088/1361-6455/aa9c99} {\bibfield  {journal} {\bibinfo
  {journal} {J. Phys. B: At. Mol. Opt. Phys.}\ }\textbf {\bibinfo {volume}
  {51}},\ \bibinfo {pages} {034005} (\bibinfo {year} {2018})}\BibitemShut
  {NoStop}%
\bibitem [{\citenamefont {Reitz}\ \emph {et~al.}(2019)\citenamefont {Reitz},
  \citenamefont {Sommer},\ and\ \citenamefont {Genes}}]{19ReSoGe}%
  \BibitemOpen
  \bibfield  {author} {\bibinfo {author} {\bibfnamefont {M.}~\bibnamefont
  {Reitz}}, \bibinfo {author} {\bibfnamefont {C.}~\bibnamefont {Sommer}},\ and\
  \bibinfo {author} {\bibfnamefont {C.}~\bibnamefont {Genes}},\ }\href
  {https://doi.org/10.1103/PhysRevLett.122.203602} {\bibfield  {journal}
  {\bibinfo  {journal} {Phys. Rev. Lett.}\ }\textbf {\bibinfo {volume} {122}},\
  \bibinfo {pages} {203602} (\bibinfo {year} {2019})}\BibitemShut {NoStop}%
\bibitem [{\citenamefont {Kansanen}\ \emph {et~al.}(2019)\citenamefont
  {Kansanen}, \citenamefont {Asikainen}, \citenamefont {Toppari}, \citenamefont
  {Groenhof},\ and\ \citenamefont {Heikkil\"a}}]{19KaAsTo}%
  \BibitemOpen
  \bibfield  {author} {\bibinfo {author} {\bibfnamefont {K.~S.~U.}\
  \bibnamefont {Kansanen}}, \bibinfo {author} {\bibfnamefont {A.}~\bibnamefont
  {Asikainen}}, \bibinfo {author} {\bibfnamefont {J.~J.}\ \bibnamefont
  {Toppari}}, \bibinfo {author} {\bibfnamefont {G.}~\bibnamefont {Groenhof}},\
  and\ \bibinfo {author} {\bibfnamefont {T.~T.}\ \bibnamefont {Heikkil\"a}},\
  }\href {https://doi.org/10.1103/PhysRevB.100.245426} {\bibfield  {journal}
  {\bibinfo  {journal} {Phys. Rev. B}\ }\textbf {\bibinfo {volume} {100}},\
  \bibinfo {pages} {245426} (\bibinfo {year} {2019})}\BibitemShut {NoStop}%
\bibitem [{\citenamefont {Groenhof}\ \emph {et~al.}(2019)\citenamefont
  {Groenhof}, \citenamefont {Climent}, \citenamefont {Feist}, \citenamefont
  {Morozov},\ and\ \citenamefont {Toppari}}]{19GrClFe}%
  \BibitemOpen
  \bibfield  {author} {\bibinfo {author} {\bibfnamefont {G.}~\bibnamefont
  {Groenhof}}, \bibinfo {author} {\bibfnamefont {C.}~\bibnamefont {Climent}},
  \bibinfo {author} {\bibfnamefont {J.}~\bibnamefont {Feist}}, \bibinfo
  {author} {\bibfnamefont {D.}~\bibnamefont {Morozov}},\ and\ \bibinfo {author}
  {\bibfnamefont {J.~J.}\ \bibnamefont {Toppari}},\ }\href
  {https://doi.org/10.1021/acs.jpclett.9b02192} {\bibfield  {journal} {\bibinfo
   {journal} {J. Phys. Chem. Lett.}\ }\textbf {\bibinfo {volume} {10}},\
  \bibinfo {pages} {5476} (\bibinfo {year} {2019})}\BibitemShut {NoStop}%
\bibitem [{\citenamefont {Ulusoy}\ \emph {et~al.}(2019)\citenamefont {Ulusoy},
  \citenamefont {Gomez},\ and\ \citenamefont {Vendrell}}]{19UlGoVe}%
  \BibitemOpen
  \bibfield  {author} {\bibinfo {author} {\bibfnamefont {I.}~\bibnamefont
  {Ulusoy}}, \bibinfo {author} {\bibfnamefont {J.}~\bibnamefont {Gomez}},\ and\
  \bibinfo {author} {\bibfnamefont {O.}~\bibnamefont {Vendrell}},\ }\href
  {https://doi.org/10.1021/acs.jpca.9b07404} {\bibfield  {journal} {\bibinfo
  {journal} {J. Phys. Chem. A}\ }\textbf {\bibinfo {volume} {123}},\ \bibinfo
  {pages} {8832} (\bibinfo {year} {2019})}\BibitemShut {NoStop}%
\bibitem [{\citenamefont {Csehi}\ \emph
  {et~al.}(2019{\natexlab{a}})\citenamefont {Csehi}, \citenamefont {Vib\'ok},
  \citenamefont {Hal\'asz},\ and\ \citenamefont {Kowalewski}}]{19CsViHa}%
  \BibitemOpen
  \bibfield  {author} {\bibinfo {author} {\bibfnamefont {A.}~\bibnamefont
  {Csehi}}, \bibinfo {author} {\bibfnamefont {A.}~\bibnamefont {Vib\'ok}},
  \bibinfo {author} {\bibfnamefont {G.~J.}\ \bibnamefont {Hal\'asz}},\ and\
  \bibinfo {author} {\bibfnamefont {M.}~\bibnamefont {Kowalewski}},\ }\href
  {https://doi.org/10.1103/PhysRevA.100.053421} {\bibfield  {journal} {\bibinfo
   {journal} {Phys. Rev. A}\ }\textbf {\bibinfo {volume} {100}},\ \bibinfo
  {pages} {053421} (\bibinfo {year} {2019}{\natexlab{a}})}\BibitemShut
  {NoStop}%
\bibitem [{\citenamefont {Csehi}\ \emph
  {et~al.}(2019{\natexlab{b}})\citenamefont {Csehi}, \citenamefont
  {Kowalewski}, \citenamefont {Hal\'asz},\ and\ \citenamefont
  {Vib\'ok}}]{19CsKoHa}%
  \BibitemOpen
  \bibfield  {author} {\bibinfo {author} {\bibfnamefont {A.}~\bibnamefont
  {Csehi}}, \bibinfo {author} {\bibfnamefont {M.}~\bibnamefont {Kowalewski}},
  \bibinfo {author} {\bibfnamefont {G.~J.}\ \bibnamefont {Hal\'asz}},\ and\
  \bibinfo {author} {\bibfnamefont {{\'A}.}~\bibnamefont {Vib\'ok}},\ }\href
  {https://doi.org/10.1088/1367-2630/ab3fcc} {\bibfield  {journal} {\bibinfo
  {journal} {New J. Phys.}\ }\textbf {\bibinfo {volume} {21}},\ \bibinfo
  {pages} {093040} (\bibinfo {year} {2019}{\natexlab{b}})}\BibitemShut
  {NoStop}%
\bibitem [{\citenamefont {Triana}\ and\ \citenamefont
  {Sanz-Vicario}(2019)}]{19TrSa}%
  \BibitemOpen
  \bibfield  {author} {\bibinfo {author} {\bibfnamefont {J.}~\bibnamefont
  {Triana}}\ and\ \bibinfo {author} {\bibfnamefont {J.}~\bibnamefont
  {Sanz-Vicario}},\ }\href {https://doi.org/10.1103/PhysRevLett.122.063603}
  {\bibfield  {journal} {\bibinfo  {journal} {Phys. Rev. Lett.}\ }\textbf
  {\bibinfo {volume} {122}},\ \bibinfo {pages} {063603} (\bibinfo {year}
  {2019})}\BibitemShut {NoStop}%
\bibitem [{\citenamefont {Gu}\ and\ \citenamefont
  {Mukamel}(2020{\natexlab{a}})}]{20GuMu}%
  \BibitemOpen
  \bibfield  {author} {\bibinfo {author} {\bibfnamefont {B.}~\bibnamefont
  {Gu}}\ and\ \bibinfo {author} {\bibfnamefont {S.}~\bibnamefont {Mukamel}},\
  }\href {https://doi.org/10.1039/c9sc04992d} {\bibfield  {journal} {\bibinfo
  {journal} {Chem. Sci.}\ }\textbf {\bibinfo {volume} {11}},\ \bibinfo {pages}
  {1290} (\bibinfo {year} {2020}{\natexlab{a}})}\BibitemShut {NoStop}%
\bibitem [{\citenamefont {Gu}\ and\ \citenamefont
  {Mukamel}(2020{\natexlab{b}})}]{20GuMu_2}%
  \BibitemOpen
  \bibfield  {author} {\bibinfo {author} {\bibfnamefont {B.}~\bibnamefont
  {Gu}}\ and\ \bibinfo {author} {\bibfnamefont {S.}~\bibnamefont {Mukamel}},\
  }\href {https://doi.org/10.1021/acs.jpclett.0c00381} {\bibfield  {journal}
  {\bibinfo  {journal} {J. Phys. Chem. Lett.}\ }\textbf {\bibinfo {volume}
  {11}},\ \bibinfo {pages} {5555} (\bibinfo {year}
  {2020}{\natexlab{b}})}\BibitemShut {NoStop}%
\bibitem [{\citenamefont {Mandal}\ \emph
  {et~al.}(2020{\natexlab{a}})\citenamefont {Mandal}, \citenamefont
  {Montillo~Vega},\ and\ \citenamefont {Huo}}]{20MaMoHu}%
  \BibitemOpen
  \bibfield  {author} {\bibinfo {author} {\bibfnamefont {A.}~\bibnamefont
  {Mandal}}, \bibinfo {author} {\bibfnamefont {S.}~\bibnamefont
  {Montillo~Vega}},\ and\ \bibinfo {author} {\bibfnamefont {P.}~\bibnamefont
  {Huo}},\ }\href {https://doi.org/10.1021/acs.jpclett.0c02399} {\bibfield
  {journal} {\bibinfo  {journal} {J. Phys. Chem. Lett.}\ }\textbf {\bibinfo
  {volume} {11}},\ \bibinfo {pages} {9215} (\bibinfo {year}
  {2020}{\natexlab{a}})}\BibitemShut {NoStop}%
\bibitem [{\citenamefont {Taylor}\ \emph {et~al.}(2020)\citenamefont {Taylor},
  \citenamefont {Mandal}, \citenamefont {Zhou},\ and\ \citenamefont
  {Huo}}]{20TaMaZh}%
  \BibitemOpen
  \bibfield  {author} {\bibinfo {author} {\bibfnamefont {M.~A.~D.}\
  \bibnamefont {Taylor}}, \bibinfo {author} {\bibfnamefont {A.}~\bibnamefont
  {Mandal}}, \bibinfo {author} {\bibfnamefont {W.}~\bibnamefont {Zhou}},\ and\
  \bibinfo {author} {\bibfnamefont {P.}~\bibnamefont {Huo}},\ }\href
  {https://doi.org/10.1103/PhysRevLett.125.123602} {\bibfield  {journal}
  {\bibinfo  {journal} {Phys. Rev. Lett.}\ }\textbf {\bibinfo {volume} {125}},\
  \bibinfo {pages} {123602} (\bibinfo {year} {2020})}\BibitemShut {NoStop}%
\bibitem [{\citenamefont {Felicetti}\ \emph {et~al.}(2020)\citenamefont
  {Felicetti}, \citenamefont {Fregoni}, \citenamefont {Schnappinger},
  \citenamefont {Reiter}, \citenamefont {De~Vivie-Riedle},\ and\ \citenamefont
  {Feist}}]{20FeFrSc}%
  \BibitemOpen
  \bibfield  {author} {\bibinfo {author} {\bibfnamefont {S.}~\bibnamefont
  {Felicetti}}, \bibinfo {author} {\bibfnamefont {J.}~\bibnamefont {Fregoni}},
  \bibinfo {author} {\bibfnamefont {T.}~\bibnamefont {Schnappinger}}, \bibinfo
  {author} {\bibfnamefont {S.}~\bibnamefont {Reiter}}, \bibinfo {author}
  {\bibfnamefont {R.}~\bibnamefont {De~Vivie-Riedle}},\ and\ \bibinfo {author}
  {\bibfnamefont {J.}~\bibnamefont {Feist}},\ }\href
  {https://doi.org/10.1021/acs.jpclett.0c02236} {\bibfield  {journal} {\bibinfo
   {journal} {J. Phys. Chem. Lett.}\ }\textbf {\bibinfo {volume} {11}},\
  \bibinfo {pages} {8810} (\bibinfo {year} {2020})}\BibitemShut {NoStop}%
\bibitem [{\citenamefont {Ulusoy}\ and\ \citenamefont
  {Vendrell}(2020)}]{20UlVe}%
  \BibitemOpen
  \bibfield  {author} {\bibinfo {author} {\bibfnamefont {I.~S.}\ \bibnamefont
  {Ulusoy}}\ and\ \bibinfo {author} {\bibfnamefont {O.}~\bibnamefont
  {Vendrell}},\ }\href {https://doi.org/10.1063/5.0011556} {\bibfield
  {journal} {\bibinfo  {journal} {J. Chem. Phys.}\ }\textbf {\bibinfo {volume}
  {153}},\ \bibinfo {pages} {044108} (\bibinfo {year} {2020})}\BibitemShut
  {NoStop}%
\bibitem [{\citenamefont {Ulusoy}\ \emph {et~al.}(2020)\citenamefont {Ulusoy},
  \citenamefont {Gomez},\ and\ \citenamefont {Vendrell}}]{20UlGoVE}%
  \BibitemOpen
  \bibfield  {author} {\bibinfo {author} {\bibfnamefont {I.~S.}\ \bibnamefont
  {Ulusoy}}, \bibinfo {author} {\bibfnamefont {J.~A.}\ \bibnamefont {Gomez}},\
  and\ \bibinfo {author} {\bibfnamefont {O.}~\bibnamefont {Vendrell}},\ }\href
  {https://doi.org/10.1063/5.0034786} {\bibfield  {journal} {\bibinfo
  {journal} {J. Chem. Phys.}\ }\textbf {\bibinfo {volume} {153}},\ \bibinfo
  {pages} {244107} (\bibinfo {year} {2020})}\BibitemShut {NoStop}%
\bibitem [{\citenamefont {Sch\"afer}\ \emph {et~al.}(2020)\citenamefont
  {Sch\"afer}, \citenamefont {Ruggenthaler}, \citenamefont {Rokaj},\ and\
  \citenamefont {Rubio}}]{20ScRuRo}%
  \BibitemOpen
  \bibfield  {author} {\bibinfo {author} {\bibfnamefont {C.}~\bibnamefont
  {Sch\"afer}}, \bibinfo {author} {\bibfnamefont {M.}~\bibnamefont
  {Ruggenthaler}}, \bibinfo {author} {\bibfnamefont {V.}~\bibnamefont
  {Rokaj}},\ and\ \bibinfo {author} {\bibfnamefont {A.}~\bibnamefont {Rubio}},\
  }\href {https://doi.org/10.1021/acsphotonics.9b01649} {\bibfield  {journal}
  {\bibinfo  {journal} {ACS Photonics}\ }\textbf {\bibinfo {volume} {7}},\
  \bibinfo {pages} {975} (\bibinfo {year} {2020})}\BibitemShut {NoStop}%
\bibitem [{\citenamefont {Sidler}\ \emph {et~al.}(2021)\citenamefont {Sidler},
  \citenamefont {Sch\"afer}, \citenamefont {Ruggenthaler},\ and\ \citenamefont
  {Rubio}}]{21SiScRu}%
  \BibitemOpen
  \bibfield  {author} {\bibinfo {author} {\bibfnamefont {D.}~\bibnamefont
  {Sidler}}, \bibinfo {author} {\bibfnamefont {C.}~\bibnamefont {Sch\"afer}},
  \bibinfo {author} {\bibfnamefont {M.}~\bibnamefont {Ruggenthaler}},\ and\
  \bibinfo {author} {\bibfnamefont {A.}~\bibnamefont {Rubio}},\ }\href
  {https://doi.org/10.1021/acs.jpclett.0c03436} {\bibfield  {journal} {\bibinfo
   {journal} {J. Phys. Chem. Lett.}\ }\textbf {\bibinfo {volume} {12}},\
  \bibinfo {pages} {508} (\bibinfo {year} {2021})}\BibitemShut {NoStop}%
\bibitem [{\citenamefont {Haugland}\ \emph {et~al.}(2021)\citenamefont
  {Haugland}, \citenamefont {Sch\"afer}, \citenamefont {Ronca}, \citenamefont
  {Rubio},\ and\ \citenamefont {Koch}}]{21HaScRo}%
  \BibitemOpen
  \bibfield  {author} {\bibinfo {author} {\bibfnamefont {T.~S.}\ \bibnamefont
  {Haugland}}, \bibinfo {author} {\bibfnamefont {C.}~\bibnamefont {Sch\"afer}},
  \bibinfo {author} {\bibfnamefont {E.}~\bibnamefont {Ronca}}, \bibinfo
  {author} {\bibfnamefont {A.}~\bibnamefont {Rubio}},\ and\ \bibinfo {author}
  {\bibfnamefont {H.}~\bibnamefont {Koch}},\ }\href
  {https://doi.org/10.1063/5.0039256} {\bibfield  {journal} {\bibinfo
  {journal} {J. Chem. Phys.}\ }\textbf {\bibinfo {volume} {154}},\ \bibinfo
  {pages} {094113} (\bibinfo {year} {2021})}\BibitemShut {NoStop}%
\bibitem [{\citenamefont {Sommer}\ \emph {et~al.}(2021)\citenamefont {Sommer},
  \citenamefont {Reitz}, \citenamefont {Mineo},\ and\ \citenamefont
  {Genes}}]{21SoReMi}%
  \BibitemOpen
  \bibfield  {author} {\bibinfo {author} {\bibfnamefont {C.}~\bibnamefont
  {Sommer}}, \bibinfo {author} {\bibfnamefont {M.}~\bibnamefont {Reitz}},
  \bibinfo {author} {\bibfnamefont {F.}~\bibnamefont {Mineo}},\ and\ \bibinfo
  {author} {\bibfnamefont {C.}~\bibnamefont {Genes}},\ }\href
  {https://doi.org/10.1103/PhysRevResearch.3.033141} {\bibfield  {journal}
  {\bibinfo  {journal} {Phys. Rev. Res.}\ }\textbf {\bibinfo {volume} {3}},\
  \bibinfo {pages} {033141} (\bibinfo {year} {2021})}\BibitemShut {NoStop}%
\bibitem [{\citenamefont {Silva}\ \emph {et~al.}(2020)\citenamefont {Silva},
  \citenamefont {Pino}, \citenamefont {Garc\'ia-Vidal},\ and\ \citenamefont
  {Feist}}]{20SiPiGa}%
  \BibitemOpen
  \bibfield  {author} {\bibinfo {author} {\bibfnamefont {R.}~\bibnamefont
  {Silva}}, \bibinfo {author} {\bibfnamefont {J.}~\bibnamefont {Pino}},
  \bibinfo {author} {\bibfnamefont {F.}~\bibnamefont {Garc\'ia-Vidal}},\ and\
  \bibinfo {author} {\bibfnamefont {J.}~\bibnamefont {Feist}},\ }\href
  {https://doi.org/10.1038/s41467-020-15196-x} {\bibfield  {journal} {\bibinfo
  {journal} {Nat. Commun.}\ }\textbf {\bibinfo {volume} {11}},\ \bibinfo
  {pages} {1423} (\bibinfo {year} {2020})}\BibitemShut {NoStop}%
\bibitem [{\citenamefont {Li}\ \emph {et~al.}(2020{\natexlab{a}})\citenamefont
  {Li}, \citenamefont {Subotnik},\ and\ \citenamefont {Nitzan}}]{20LiSuNi}%
  \BibitemOpen
  \bibfield  {author} {\bibinfo {author} {\bibfnamefont {T.~E.}\ \bibnamefont
  {Li}}, \bibinfo {author} {\bibfnamefont {J.~E.}\ \bibnamefont {Subotnik}},\
  and\ \bibinfo {author} {\bibfnamefont {A.}~\bibnamefont {Nitzan}},\ }\href
  {https://doi.org/10.1073/pnas.2009272117} {\bibfield  {journal} {\bibinfo
  {journal} {Proc. Natl. Acad. Sci. U.S.A.}\ }\textbf {\bibinfo {volume}
  {117}},\ \bibinfo {pages} {18324} (\bibinfo {year}
  {2020}{\natexlab{a}})}\BibitemShut {NoStop}%
\bibitem [{\citenamefont {Li}\ \emph {et~al.}(2020{\natexlab{b}})\citenamefont
  {Li}, \citenamefont {Chen}, \citenamefont {Nitzan},\ and\ \citenamefont
  {Subotnik}}]{20LiChNi}%
  \BibitemOpen
  \bibfield  {author} {\bibinfo {author} {\bibfnamefont {T.~E.}\ \bibnamefont
  {Li}}, \bibinfo {author} {\bibfnamefont {H.-T.}\ \bibnamefont {Chen}},
  \bibinfo {author} {\bibfnamefont {A.}~\bibnamefont {Nitzan}},\ and\ \bibinfo
  {author} {\bibfnamefont {J.~E.}\ \bibnamefont {Subotnik}},\ }\href
  {https://doi.org/10.1103/PhysRevA.101.033831} {\bibfield  {journal} {\bibinfo
   {journal} {Phys. Rev. A}\ }\textbf {\bibinfo {volume} {101}},\ \bibinfo
  {pages} {033831} (\bibinfo {year} {2020}{\natexlab{b}})}\BibitemShut
  {NoStop}%
\bibitem [{\citenamefont {Davidsson}\ and\ \citenamefont
  {Kowalewski}(2020)}]{20DaKo}%
  \BibitemOpen
  \bibfield  {author} {\bibinfo {author} {\bibfnamefont {E.}~\bibnamefont
  {Davidsson}}\ and\ \bibinfo {author} {\bibfnamefont {M.}~\bibnamefont
  {Kowalewski}},\ }\href {https://doi.org/10.1063/5.0033773} {\bibfield
  {journal} {\bibinfo  {journal} {J. Chem. Phys.}\ }\textbf {\bibinfo {volume}
  {153}},\ \bibinfo {pages} {234304} (\bibinfo {year} {2020})}\BibitemShut
  {NoStop}%
\bibitem [{\citenamefont {Spano}(2020)}]{20Spano}%
  \BibitemOpen
  \bibfield  {author} {\bibinfo {author} {\bibfnamefont {F.~C.}\ \bibnamefont
  {Spano}},\ }\href {https://doi.org/10.1063/5.0002164} {\bibfield  {journal}
  {\bibinfo  {journal} {J. Chem. Phys.}\ }\textbf {\bibinfo {volume} {152}},\
  \bibinfo {pages} {204113} (\bibinfo {year} {2020})}\BibitemShut {NoStop}%
\bibitem [{\citenamefont {Triana}\ and\ \citenamefont
  {Sanz-Vicario}(2021)}]{21TrSa}%
  \BibitemOpen
  \bibfield  {author} {\bibinfo {author} {\bibfnamefont {J.}~\bibnamefont
  {Triana}}\ and\ \bibinfo {author} {\bibfnamefont {J.}~\bibnamefont
  {Sanz-Vicario}},\ }\href {https://doi.org/10.1063/5.0037995} {\bibfield
  {journal} {\bibinfo  {journal} {J. Chem. Phys.}\ }\textbf {\bibinfo {volume}
  {154}},\ \bibinfo {pages} {094120} (\bibinfo {year} {2021})}\BibitemShut
  {NoStop}%
\bibitem [{\citenamefont {Gudem}\ and\ \citenamefont
  {Kowalewski}(2021)}]{21GuKo}%
  \BibitemOpen
  \bibfield  {author} {\bibinfo {author} {\bibfnamefont {M.}~\bibnamefont
  {Gudem}}\ and\ \bibinfo {author} {\bibfnamefont {M.}~\bibnamefont
  {Kowalewski}},\ }\href {https://doi.org/10.1021/acs.jpca.0c09252} {\bibfield
  {journal} {\bibinfo  {journal} {J. Phys. Chem. A}\ }\textbf {\bibinfo
  {volume} {125}},\ \bibinfo {pages} {1142} (\bibinfo {year}
  {2021})}\BibitemShut {NoStop}%
\bibitem [{\citenamefont {Li}\ \emph {et~al.}(2021{\natexlab{a}})\citenamefont
  {Li}, \citenamefont {Nitzan},\ and\ \citenamefont {Subotnik}}]{21LiNiSu}%
  \BibitemOpen
  \bibfield  {author} {\bibinfo {author} {\bibfnamefont {T.~E.}\ \bibnamefont
  {Li}}, \bibinfo {author} {\bibfnamefont {A.}~\bibnamefont {Nitzan}},\ and\
  \bibinfo {author} {\bibfnamefont {J.~E.}\ \bibnamefont {Subotnik}},\ }\href
  {https://doi.org/https://doi.org/10.1002/anie.202103920} {\bibfield
  {journal} {\bibinfo  {journal} {Angew. Chem. Int. Ed.}\ }\textbf {\bibinfo
  {volume} {60}},\ \bibinfo {pages} {15533} (\bibinfo {year}
  {2021}{\natexlab{a}})}\BibitemShut {NoStop}%
\bibitem [{\citenamefont {Tichauer}\ \emph {et~al.}(2021)\citenamefont
  {Tichauer}, \citenamefont {Feist},\ and\ \citenamefont
  {Groenhof}}]{21TiFeGr}%
  \BibitemOpen
  \bibfield  {author} {\bibinfo {author} {\bibfnamefont {R.~H.}\ \bibnamefont
  {Tichauer}}, \bibinfo {author} {\bibfnamefont {J.}~\bibnamefont {Feist}},\
  and\ \bibinfo {author} {\bibfnamefont {G.}~\bibnamefont {Groenhof}},\ }\href
  {https://doi.org/10.1063/5.0037868} {\bibfield  {journal} {\bibinfo
  {journal} {J. Chem. Phys.}\ }\textbf {\bibinfo {volume} {154}},\ \bibinfo
  {pages} {104112} (\bibinfo {year} {2021})}\BibitemShut {NoStop}%
\bibitem [{\citenamefont {Cederbaum}\ and\ \citenamefont
  {Kuleff}(2021)}]{21CeKu}%
  \BibitemOpen
  \bibfield  {author} {\bibinfo {author} {\bibfnamefont {L.~S.}\ \bibnamefont
  {Cederbaum}}\ and\ \bibinfo {author} {\bibfnamefont {A.~I.}\ \bibnamefont
  {Kuleff}},\ }\href {https://doi.org/10.1038/s41467-021-24221-6} {\bibfield
  {journal} {\bibinfo  {journal} {Nat. Commun.}\ }\textbf {\bibinfo {volume}
  {12}},\ \bibinfo {pages} {4083} (\bibinfo {year} {2021})}\BibitemShut
  {NoStop}%
\bibitem [{\citenamefont {Cederbaum}(2021)}]{21Cederbaum}%
  \BibitemOpen
  \bibfield  {author} {\bibinfo {author} {\bibfnamefont {L.~S.}\ \bibnamefont
  {Cederbaum}},\ }\href {https://doi.org/10.1021/acs.jpclett.1c01570}
  {\bibfield  {journal} {\bibinfo  {journal} {J. Phys. Chem. Lett.}\ }\textbf
  {\bibinfo {volume} {12}},\ \bibinfo {pages} {6056} (\bibinfo {year}
  {2021})}\BibitemShut {NoStop}%
\bibitem [{\citenamefont {Szidarovszky}\ \emph {et~al.}(2021)\citenamefont
  {Szidarovszky}, \citenamefont {Badank\'o}, \citenamefont {Hal\'asz},\ and\
  \citenamefont {Vib\'ok}}]{21SzBaHa}%
  \BibitemOpen
  \bibfield  {author} {\bibinfo {author} {\bibfnamefont {T.}~\bibnamefont
  {Szidarovszky}}, \bibinfo {author} {\bibfnamefont {P.}~\bibnamefont
  {Badank\'o}}, \bibinfo {author} {\bibfnamefont {G.~J.}\ \bibnamefont
  {Hal\'asz}},\ and\ \bibinfo {author} {\bibfnamefont {{\'A}.}~\bibnamefont
  {Vib\'ok}},\ }\href {https://doi.org/10.1063/5.0033338} {\bibfield  {journal}
  {\bibinfo  {journal} {J. Chem. Phys.}\ }\textbf {\bibinfo {volume} {154}},\
  \bibinfo {pages} {064305} (\bibinfo {year} {2021})}\BibitemShut {NoStop}%
\bibitem [{\citenamefont {Silva}\ and\ \citenamefont {Feist}(2022)}]{22SiFe}%
  \BibitemOpen
  \bibfield  {author} {\bibinfo {author} {\bibfnamefont {R.~E.~F.}\
  \bibnamefont {Silva}}\ and\ \bibinfo {author} {\bibfnamefont
  {J.}~\bibnamefont {Feist}},\ }\href
  {https://doi.org/10.1103/PhysRevA.105.043704} {\bibfield  {journal} {\bibinfo
   {journal} {Phys. Rev. A}\ }\textbf {\bibinfo {volume} {105}},\ \bibinfo
  {pages} {043704} (\bibinfo {year} {2022})}\BibitemShut {NoStop}%
\bibitem [{\citenamefont {Cederbaum}(2022)}]{22Cederbaum}%
  \BibitemOpen
  \bibfield  {author} {\bibinfo {author} {\bibfnamefont {L.~S.}\ \bibnamefont
  {Cederbaum}},\ }\href {https://doi.org/10.1063/5.0090047} {\bibfield
  {journal} {\bibinfo  {journal} {J. Chem. Phys.}\ }\textbf {\bibinfo {volume}
  {156}},\ \bibinfo {pages} {184102} (\bibinfo {year} {2022})}\BibitemShut
  {NoStop}%
\bibitem [{\citenamefont {Li}\ \emph {et~al.}(2022{\natexlab{a}})\citenamefont
  {Li}, \citenamefont {Nitzan}, \citenamefont {Hammes-Schiffer},\ and\
  \citenamefont {Subotnik}}]{22LiNiHa}%
  \BibitemOpen
  \bibfield  {author} {\bibinfo {author} {\bibfnamefont {T.~E.}\ \bibnamefont
  {Li}}, \bibinfo {author} {\bibfnamefont {A.}~\bibnamefont {Nitzan}}, \bibinfo
  {author} {\bibfnamefont {S.}~\bibnamefont {Hammes-Schiffer}},\ and\ \bibinfo
  {author} {\bibfnamefont {J.~E.}\ \bibnamefont {Subotnik}},\ }\href
  {https://doi.org/10.1021/acs.jpclett.2c00613} {\bibfield  {journal} {\bibinfo
   {journal} {J. Phys. Chem. Lett.}\ }\textbf {\bibinfo {volume} {13}},\
  \bibinfo {pages} {3890} (\bibinfo {year} {2022}{\natexlab{a}})}\BibitemShut
  {NoStop}%
\bibitem [{\citenamefont {Csehi}\ \emph {et~al.}(2022)\citenamefont {Csehi},
  \citenamefont {Vendrell}, \citenamefont {Hal{\'{a}}sz},\ and\ \citenamefont
  {Vib{\'{o}}k}}]{22CsVeHa}%
  \BibitemOpen
  \bibfield  {author} {\bibinfo {author} {\bibfnamefont {A.}~\bibnamefont
  {Csehi}}, \bibinfo {author} {\bibfnamefont {O.}~\bibnamefont {Vendrell}},
  \bibinfo {author} {\bibfnamefont {G.~J.}\ \bibnamefont {Hal{\'{a}}sz}},\ and\
  \bibinfo {author} {\bibfnamefont {{\'{A}}.}~\bibnamefont {Vib{\'{o}}k}},\
  }\href {https://doi.org/10.1088/1367-2630/ac7df7} {\bibfield  {journal}
  {\bibinfo  {journal} {New J. Phys.}\ }\textbf {\bibinfo {volume} {24}},\
  \bibinfo {pages} {073022} (\bibinfo {year} {2022})}\BibitemShut {NoStop}%
\bibitem [{\citenamefont {F{\'{a}}bri}\ \emph {et~al.}(2022)\citenamefont
  {F{\'{a}}bri}, \citenamefont {Hal{\'{a}}sz},\ and\ \citenamefont
  {Vib{\'{o}}k}}]{22FaHaVi}%
  \BibitemOpen
  \bibfield  {author} {\bibinfo {author} {\bibfnamefont {C.}~\bibnamefont
  {F{\'{a}}bri}}, \bibinfo {author} {\bibfnamefont {G.~J.}\ \bibnamefont
  {Hal{\'{a}}sz}},\ and\ \bibinfo {author} {\bibfnamefont
  {{\'{A}}.}~\bibnamefont {Vib{\'{o}}k}},\ }\href
  {https://doi.org/10.1021/acs.jpclett.1c03465} {\bibfield  {journal} {\bibinfo
   {journal} {J. Phys. Chem. Lett.}\ }\textbf {\bibinfo {volume} {13}},\
  \bibinfo {pages} {1172} (\bibinfo {year} {2022})}\BibitemShut {NoStop}%
\bibitem [{\citenamefont {F{\'{a}}bri}\ \emph {et~al.}(2023)\citenamefont
  {F{\'{a}}bri}, \citenamefont {Hal{\'{a}}sz}, \citenamefont {Cederbaum},\ and\
  \citenamefont {Vib{\'{o}}k}}]{23FaHaCe}%
  \BibitemOpen
  \bibfield  {author} {\bibinfo {author} {\bibfnamefont {C.}~\bibnamefont
  {F{\'{a}}bri}}, \bibinfo {author} {\bibfnamefont {G.~J.}\ \bibnamefont
  {Hal{\'{a}}sz}}, \bibinfo {author} {\bibfnamefont {L.~S.}\ \bibnamefont
  {Cederbaum}},\ and\ \bibinfo {author} {\bibfnamefont {{\'{A}}.}~\bibnamefont
  {Vib{\'{o}}k}},\ }\href {https://doi.org/10.1039/d2cc04222c} {\bibfield
  {journal} {\bibinfo  {journal} {Chem. Commun.}\ }\textbf {\bibinfo {volume}
  {58}},\ \bibinfo {pages} {12612} (\bibinfo {year} {2023})}\BibitemShut
  {NoStop}%
\bibitem [{\citenamefont {P\'erez-S\'anchez}\ \emph {et~al.}(2023)\citenamefont
  {P\'erez-S\'anchez}, \citenamefont {Koner}, \citenamefont {Stern},\ and\
  \citenamefont {Yuen-Zhou}}]{23PeKoSt}%
  \BibitemOpen
  \bibfield  {author} {\bibinfo {author} {\bibfnamefont {J.~B.}\ \bibnamefont
  {P\'erez-S\'anchez}}, \bibinfo {author} {\bibfnamefont {A.}~\bibnamefont
  {Koner}}, \bibinfo {author} {\bibfnamefont {N.~P.}\ \bibnamefont {Stern}},\
  and\ \bibinfo {author} {\bibfnamefont {J.}~\bibnamefont {Yuen-Zhou}},\ }\href
  {https://doi.org/10.1073/pnas.2219223120} {\bibfield  {journal} {\bibinfo
  {journal} {Proc. Natl. Acad. Sci. U.S.A.}\ }\textbf {\bibinfo {volume}
  {120}},\ \bibinfo {pages} {e2219223120} (\bibinfo {year} {2023})}\BibitemShut
  {NoStop}%
\bibitem [{\citenamefont {Nandipati}\ and\ \citenamefont
  {Vendrell}(2023)}]{23NaVe}%
  \BibitemOpen
  \bibfield  {author} {\bibinfo {author} {\bibfnamefont {K.~R.}\ \bibnamefont
  {Nandipati}}\ and\ \bibinfo {author} {\bibfnamefont {O.}~\bibnamefont
  {Vendrell}},\ }\href {https://doi.org/10.1103/PhysRevA.107.L061101}
  {\bibfield  {journal} {\bibinfo  {journal} {Phys. Rev. A}\ }\textbf {\bibinfo
  {volume} {107}},\ \bibinfo {pages} {L061101} (\bibinfo {year}
  {2023})}\BibitemShut {NoStop}%
\bibitem [{\citenamefont {Schnappinger}\ and\ \citenamefont
  {Kowalewski}(2023)}]{23ScKo}%
  \BibitemOpen
  \bibfield  {author} {\bibinfo {author} {\bibfnamefont {T.}~\bibnamefont
  {Schnappinger}}\ and\ \bibinfo {author} {\bibfnamefont {M.}~\bibnamefont
  {Kowalewski}},\ }\href {https://doi.org/10.1021/acs.jctc.2c01154} {\bibfield
  {journal} {\bibinfo  {journal} {J. Chem. Theory Comput.}\ }\textbf {\bibinfo
  {volume} {19}},\ \bibinfo {pages} {460} (\bibinfo {year} {2023})}\BibitemShut
  {NoStop}%
\bibitem [{\citenamefont {Szidarovszky}(2023{\natexlab{a}})}]{23Szidarovszky}%
  \BibitemOpen
  \bibfield  {author} {\bibinfo {author} {\bibfnamefont {T.}~\bibnamefont
  {Szidarovszky}},\ }\href {https://doi.org/10.1063/5.0153293} {\bibfield
  {journal} {\bibinfo  {journal} {J. Chem. Phys.}\ }\textbf {\bibinfo {volume}
  {159}},\ \bibinfo {pages} {014112} (\bibinfo {year}
  {2023}{\natexlab{a}})}\BibitemShut {NoStop}%
\bibitem [{\citenamefont {Szidarovszky}(2023{\natexlab{b}})}]{23Szidarovszky2}%
  \BibitemOpen
  \bibfield  {author} {\bibinfo {author} {\bibfnamefont {T.}~\bibnamefont
  {Szidarovszky}},\ }\href {https://doi.org/10.48550/arXiv.2307.03508}
  {\bibinfo {title} {Pauli principle in polaritonic chemistry}} (\bibinfo
  {year} {2023}{\natexlab{b}}),\ \Eprint {https://arxiv.org/abs/2307.03508}
  {arXiv:2307.03508 [quant-ph]} \BibitemShut {NoStop}%
\bibitem [{\citenamefont {Galego}\ \emph {et~al.}(2015)\citenamefont {Galego},
  \citenamefont {Garcia-Vidal},\ and\ \citenamefont {Feist}}]{15GaGaFe}%
  \BibitemOpen
  \bibfield  {author} {\bibinfo {author} {\bibfnamefont {J.}~\bibnamefont
  {Galego}}, \bibinfo {author} {\bibfnamefont {F.~J.}\ \bibnamefont
  {Garcia-Vidal}},\ and\ \bibinfo {author} {\bibfnamefont {J.}~\bibnamefont
  {Feist}},\ }\href {https://doi.org/10.1103/PhysRevX.5.041022} {\bibfield
  {journal} {\bibinfo  {journal} {Phys. Rev. X}\ }\textbf {\bibinfo {volume}
  {5}},\ \bibinfo {pages} {041022} (\bibinfo {year} {2015})}\BibitemShut
  {NoStop}%
\bibitem [{\citenamefont {Herrera}\ and\ \citenamefont {Spano}(2017)}]{17HeSp}%
  \BibitemOpen
  \bibfield  {author} {\bibinfo {author} {\bibfnamefont {F.}~\bibnamefont
  {Herrera}}\ and\ \bibinfo {author} {\bibfnamefont {F.~C.}\ \bibnamefont
  {Spano}},\ }\href {https://doi.org/10.1103/PhysRevLett.118.223601} {\bibfield
   {journal} {\bibinfo  {journal} {Phys. Rev. Lett.}\ }\textbf {\bibinfo
  {volume} {118}},\ \bibinfo {pages} {223601} (\bibinfo {year}
  {2017})}\BibitemShut {NoStop}%
\bibitem [{\citenamefont {Herrera}\ and\ \citenamefont {Spano}(2018)}]{18HeSp}%
  \BibitemOpen
  \bibfield  {author} {\bibinfo {author} {\bibfnamefont {F.}~\bibnamefont
  {Herrera}}\ and\ \bibinfo {author} {\bibfnamefont {F.~C.}\ \bibnamefont
  {Spano}},\ }\href {https://doi.org/10.1021/acsphotonics.7b00728} {\bibfield
  {journal} {\bibinfo  {journal} {ACS Photonics}\ }\textbf {\bibinfo {volume}
  {5}},\ \bibinfo {pages} {65} (\bibinfo {year} {2018})}\BibitemShut {NoStop}%
\bibitem [{\citenamefont {Xiang}\ \emph {et~al.}(2018)\citenamefont {Xiang},
  \citenamefont {Ribeiro}, \citenamefont {Dunkelberger}, \citenamefont {Wang},
  \citenamefont {Li}, \citenamefont {Simpkins}, \citenamefont {Owrutsky},
  \citenamefont {Yuen-Zhou},\ and\ \citenamefont {Xiong}}]{18XiRiDu}%
  \BibitemOpen
  \bibfield  {author} {\bibinfo {author} {\bibfnamefont {B.}~\bibnamefont
  {Xiang}}, \bibinfo {author} {\bibfnamefont {R.~F.}\ \bibnamefont {Ribeiro}},
  \bibinfo {author} {\bibfnamefont {A.~D.}\ \bibnamefont {Dunkelberger}},
  \bibinfo {author} {\bibfnamefont {J.}~\bibnamefont {Wang}}, \bibinfo {author}
  {\bibfnamefont {Y.}~\bibnamefont {Li}}, \bibinfo {author} {\bibfnamefont
  {B.~S.}\ \bibnamefont {Simpkins}}, \bibinfo {author} {\bibfnamefont {J.~C.}\
  \bibnamefont {Owrutsky}}, \bibinfo {author} {\bibfnamefont {J.}~\bibnamefont
  {Yuen-Zhou}},\ and\ \bibinfo {author} {\bibfnamefont {W.}~\bibnamefont
  {Xiong}},\ }\href@noop {} {\bibfield  {journal} {\bibinfo  {journal} {Proc.
  Natl. Acad. Sci. U.S.A.}\ }\textbf {\bibinfo {volume} {115}},\ \bibinfo
  {pages} {4845} (\bibinfo {year} {2018})}\BibitemShut {NoStop}%
\bibitem [{\citenamefont {Szidarovszky}\ \emph {et~al.}(2018)\citenamefont
  {Szidarovszky}, \citenamefont {Hal\'asz}, \citenamefont {Cs\'asz\'ar},
  \citenamefont {Cederbaum},\ and\ \citenamefont {Vib\'ok}}]{18SzHaCs}%
  \BibitemOpen
  \bibfield  {author} {\bibinfo {author} {\bibfnamefont {T.}~\bibnamefont
  {Szidarovszky}}, \bibinfo {author} {\bibfnamefont {G.}~\bibnamefont
  {Hal\'asz}}, \bibinfo {author} {\bibfnamefont {A.}~\bibnamefont
  {Cs\'asz\'ar}}, \bibinfo {author} {\bibfnamefont {L.}~\bibnamefont
  {Cederbaum}},\ and\ \bibinfo {author} {\bibfnamefont {{\'A}.}~\bibnamefont
  {Vib\'ok}},\ }\href {https://doi.org/10.1021/acs.jpclett.8b02609} {\bibfield
  {journal} {\bibinfo  {journal} {J. Phys. Chem. Lett.}\ }\textbf {\bibinfo
  {volume} {9}},\ \bibinfo {pages} {6215} (\bibinfo {year} {2018})}\BibitemShut
  {NoStop}%
\bibitem [{\citenamefont {Szidarovszky}\ \emph {et~al.}(2020)\citenamefont
  {Szidarovszky}, \citenamefont {Hal{\'{a}}sz},\ and\ \citenamefont
  {Vib{\'{o}}k}}]{20SzHaVi}%
  \BibitemOpen
  \bibfield  {author} {\bibinfo {author} {\bibfnamefont {T.}~\bibnamefont
  {Szidarovszky}}, \bibinfo {author} {\bibfnamefont {G.~J.}\ \bibnamefont
  {Hal{\'{a}}sz}},\ and\ \bibinfo {author} {\bibfnamefont
  {{\'{A}}.}~\bibnamefont {Vib{\'{o}}k}},\ }\href
  {https://doi.org/10.1088/1367-2630/ab8264} {\bibfield  {journal} {\bibinfo
  {journal} {New J. Phys.}\ }\textbf {\bibinfo {volume} {22}},\ \bibinfo
  {pages} {053001} (\bibinfo {year} {2020})}\BibitemShut {NoStop}%
\bibitem [{\citenamefont {F\'abri}\ \emph {et~al.}(2020)\citenamefont
  {F\'abri}, \citenamefont {Lasorne}, \citenamefont {Hal\'asz}, \citenamefont
  {Cederbaum},\ and\ \citenamefont {Vib\'ok}}]{20FaLaHa}%
  \BibitemOpen
  \bibfield  {author} {\bibinfo {author} {\bibfnamefont {C.}~\bibnamefont
  {F\'abri}}, \bibinfo {author} {\bibfnamefont {B.}~\bibnamefont {Lasorne}},
  \bibinfo {author} {\bibfnamefont {G.}~\bibnamefont {Hal\'asz}}, \bibinfo
  {author} {\bibfnamefont {L.}~\bibnamefont {Cederbaum}},\ and\ \bibinfo
  {author} {\bibfnamefont {{\'A}.}~\bibnamefont {Vib\'ok}},\ }\href
  {https://doi.org/10.1063/5.0035870} {\bibfield  {journal} {\bibinfo
  {journal} {J. Chem. Phys.}\ }\textbf {\bibinfo {volume} {153}},\ \bibinfo
  {pages} {234302} (\bibinfo {year} {2020})}\BibitemShut {NoStop}%
\bibitem [{\citenamefont {F\'abri}\ \emph {et~al.}(2021)\citenamefont
  {F\'abri}, \citenamefont {Hal\'asz}, \citenamefont {Cederbaum},\ and\
  \citenamefont {Vib\'ok}}]{21FaHaCe}%
  \BibitemOpen
  \bibfield  {author} {\bibinfo {author} {\bibfnamefont {C.}~\bibnamefont
  {F\'abri}}, \bibinfo {author} {\bibfnamefont {G.}~\bibnamefont {Hal\'asz}},
  \bibinfo {author} {\bibfnamefont {L.}~\bibnamefont {Cederbaum}},\ and\
  \bibinfo {author} {\bibfnamefont {{\'A}.}~\bibnamefont {Vib\'ok}},\ }\href
  {https://doi.org/10.1039/d0sc05164k} {\bibfield  {journal} {\bibinfo
  {journal} {Chem. Sci.}\ }\textbf {\bibinfo {volume} {12}},\ \bibinfo {pages}
  {1251} (\bibinfo {year} {2021})}\BibitemShut {NoStop}%
\bibitem [{\citenamefont {Farag}\ \emph {et~al.}(2021)\citenamefont {Farag},
  \citenamefont {Mandal},\ and\ \citenamefont {Huo}}]{21FaMaHu}%
  \BibitemOpen
  \bibfield  {author} {\bibinfo {author} {\bibfnamefont {M.~H.}\ \bibnamefont
  {Farag}}, \bibinfo {author} {\bibfnamefont {A.}~\bibnamefont {Mandal}},\ and\
  \bibinfo {author} {\bibfnamefont {P.}~\bibnamefont {Huo}},\ }\href
  {https://doi.org/10.1039/D1CP00943E} {\bibfield  {journal} {\bibinfo
  {journal} {Phys. Chem. Chem. Phys.}\ }\textbf {\bibinfo {volume} {23}},\
  \bibinfo {pages} {16868} (\bibinfo {year} {2021})}\BibitemShut {NoStop}%
\bibitem [{\citenamefont {Badank{\'o}}\ \emph {et~al.}(2022)\citenamefont
  {Badank{\'o}}, \citenamefont {Umarov}, \citenamefont {F{\'a}bri},
  \citenamefont {Hal{\'a}sz},\ and\ \citenamefont {Vib{\'o}k}}]{22BaUmFa}%
  \BibitemOpen
  \bibfield  {author} {\bibinfo {author} {\bibfnamefont {P.}~\bibnamefont
  {Badank{\'o}}}, \bibinfo {author} {\bibfnamefont {O.}~\bibnamefont {Umarov}},
  \bibinfo {author} {\bibfnamefont {C.}~\bibnamefont {F{\'a}bri}}, \bibinfo
  {author} {\bibfnamefont {G.}~\bibnamefont {Hal{\'a}sz}},\ and\ \bibinfo
  {author} {\bibfnamefont {{\'A}.}~\bibnamefont {Vib{\'o}k}},\ }\href
  {https://doi.org/https://doi.org/10.1002/qua.26750} {\bibfield  {journal}
  {\bibinfo  {journal} {Int. J. Quantum Chem.}\ }\textbf {\bibinfo {volume}
  {122}},\ \bibinfo {pages} {e26750} (\bibinfo {year} {2022})}\BibitemShut
  {NoStop}%
\bibitem [{\citenamefont {Galego}\ \emph {et~al.}(2016)\citenamefont {Galego},
  \citenamefont {Garcia-Vidal},\ and\ \citenamefont {Feist}}]{16GaGaFe}%
  \BibitemOpen
  \bibfield  {author} {\bibinfo {author} {\bibfnamefont {J.}~\bibnamefont
  {Galego}}, \bibinfo {author} {\bibfnamefont {F.~J.}\ \bibnamefont
  {Garcia-Vidal}},\ and\ \bibinfo {author} {\bibfnamefont {J.}~\bibnamefont
  {Feist}},\ }\href {https://doi.org/10.1038/ncomms13841} {\bibfield  {journal}
  {\bibinfo  {journal} {Nat. Commun.}\ }\textbf {\bibinfo {volume} {7}},\
  \bibinfo {pages} {13841} (\bibinfo {year} {2016})}\BibitemShut {NoStop}%
\bibitem [{\citenamefont {Herrera}\ and\ \citenamefont {Spano}(2016)}]{16HeSp}%
  \BibitemOpen
  \bibfield  {author} {\bibinfo {author} {\bibfnamefont {F.}~\bibnamefont
  {Herrera}}\ and\ \bibinfo {author} {\bibfnamefont {F.~C.}\ \bibnamefont
  {Spano}},\ }\href {https://doi.org/10.1103/PhysRevLett.116.238301} {\bibfield
   {journal} {\bibinfo  {journal} {Phys. Rev. Lett.}\ }\textbf {\bibinfo
  {volume} {116}},\ \bibinfo {pages} {238301} (\bibinfo {year}
  {2016})}\BibitemShut {NoStop}%
\bibitem [{\citenamefont {Galego}\ \emph {et~al.}(2019)\citenamefont {Galego},
  \citenamefont {Climent}, \citenamefont {Garcia-Vidal},\ and\ \citenamefont
  {Feist}}]{19GaClGa}%
  \BibitemOpen
  \bibfield  {author} {\bibinfo {author} {\bibfnamefont {J.}~\bibnamefont
  {Galego}}, \bibinfo {author} {\bibfnamefont {C.}~\bibnamefont {Climent}},
  \bibinfo {author} {\bibfnamefont {F.~J.}\ \bibnamefont {Garcia-Vidal}},\ and\
  \bibinfo {author} {\bibfnamefont {J.}~\bibnamefont {Feist}},\ }\href
  {https://doi.org/10.1103/PhysRevX.9.021057} {\bibfield  {journal} {\bibinfo
  {journal} {Phys. Rev. X}\ }\textbf {\bibinfo {volume} {9}},\ \bibinfo {pages}
  {021057} (\bibinfo {year} {2019})}\BibitemShut {NoStop}%
\bibitem [{\citenamefont {Mandal}\ and\ \citenamefont {Huo}(2019)}]{19MaHu}%
  \BibitemOpen
  \bibfield  {author} {\bibinfo {author} {\bibfnamefont {A.}~\bibnamefont
  {Mandal}}\ and\ \bibinfo {author} {\bibfnamefont {P.}~\bibnamefont {Huo}},\
  }\href {https://doi.org/10.1021/acs.jpclett.9b01599} {\bibfield  {journal}
  {\bibinfo  {journal} {J. Phys. Chem. Lett.}\ }\textbf {\bibinfo {volume}
  {10}},\ \bibinfo {pages} {5519} (\bibinfo {year} {2019})}\BibitemShut
  {NoStop}%
\bibitem [{\citenamefont {Mandal}\ \emph
  {et~al.}(2020{\natexlab{b}})\citenamefont {Mandal}, \citenamefont {Krauss},\
  and\ \citenamefont {Huo}}]{20MaKrHu}%
  \BibitemOpen
  \bibfield  {author} {\bibinfo {author} {\bibfnamefont {A.}~\bibnamefont
  {Mandal}}, \bibinfo {author} {\bibfnamefont {T.~D.}\ \bibnamefont {Krauss}},\
  and\ \bibinfo {author} {\bibfnamefont {P.}~\bibnamefont {Huo}},\ }\href
  {https://doi.org/10.1021/acs.jpcb.0c03227} {\bibfield  {journal} {\bibinfo
  {journal} {J. Phys. Chem. B}\ }\textbf {\bibinfo {volume} {124}},\ \bibinfo
  {pages} {6321} (\bibinfo {year} {2020}{\natexlab{b}})}\BibitemShut {NoStop}%
\bibitem [{\citenamefont {Li}\ \emph {et~al.}(2020{\natexlab{c}})\citenamefont
  {Li}, \citenamefont {Nitzan},\ and\ \citenamefont {Subotnik}}]{20LiNiSu}%
  \BibitemOpen
  \bibfield  {author} {\bibinfo {author} {\bibfnamefont {T.~E.}\ \bibnamefont
  {Li}}, \bibinfo {author} {\bibfnamefont {A.}~\bibnamefont {Nitzan}},\ and\
  \bibinfo {author} {\bibfnamefont {J.~E.}\ \bibnamefont {Subotnik}},\ }\href
  {https://doi.org/10.1063/5.0006472} {\bibfield  {journal} {\bibinfo
  {journal} {J. Chem. Phys.}\ }\textbf {\bibinfo {volume} {152}},\ \bibinfo
  {pages} {234107} (\bibinfo {year} {2020}{\natexlab{c}})}\BibitemShut
  {NoStop}%
\bibitem [{\citenamefont {Hoffmann}\ \emph {et~al.}(2020)\citenamefont
  {Hoffmann}, \citenamefont {Lacombe}, \citenamefont {Rubio},\ and\
  \citenamefont {Maitra}}]{20HoLaRu}%
  \BibitemOpen
  \bibfield  {author} {\bibinfo {author} {\bibfnamefont {N.~M.}\ \bibnamefont
  {Hoffmann}}, \bibinfo {author} {\bibfnamefont {L.}~\bibnamefont {Lacombe}},
  \bibinfo {author} {\bibfnamefont {A.}~\bibnamefont {Rubio}},\ and\ \bibinfo
  {author} {\bibfnamefont {N.~T.}\ \bibnamefont {Maitra}},\ }\href
  {https://doi.org/10.1063/5.0012723} {\bibfield  {journal} {\bibinfo
  {journal} {J. Chem. Phys.}\ }\textbf {\bibinfo {volume} {153}},\ \bibinfo
  {pages} {104103} (\bibinfo {year} {2020})}\BibitemShut {NoStop}%
\bibitem [{\citenamefont {Li}\ \emph {et~al.}(2021{\natexlab{b}})\citenamefont
  {Li}, \citenamefont {Mandal},\ and\ \citenamefont {Huo}}]{21LiMaHu}%
  \BibitemOpen
  \bibfield  {author} {\bibinfo {author} {\bibfnamefont {X.}~\bibnamefont
  {Li}}, \bibinfo {author} {\bibfnamefont {A.}~\bibnamefont {Mandal}},\ and\
  \bibinfo {author} {\bibfnamefont {P.}~\bibnamefont {Huo}},\ }\href
  {https://doi.org/10.1038/s41467-021-21610-9} {\bibfield  {journal} {\bibinfo
  {journal} {Nat Commun.}\ }\textbf {\bibinfo {volume} {12}},\ \bibinfo {pages}
  {1315} (\bibinfo {year} {2021}{\natexlab{b}})}\BibitemShut {NoStop}%
\bibitem [{\citenamefont {Li}\ \emph {et~al.}(2021{\natexlab{c}})\citenamefont
  {Li}, \citenamefont {Mandal},\ and\ \citenamefont {Huo}}]{21LiMaHu_2}%
  \BibitemOpen
  \bibfield  {author} {\bibinfo {author} {\bibfnamefont {X.}~\bibnamefont
  {Li}}, \bibinfo {author} {\bibfnamefont {A.}~\bibnamefont {Mandal}},\ and\
  \bibinfo {author} {\bibfnamefont {P.}~\bibnamefont {Huo}},\ }\href
  {https://doi.org/10.1021/acs.jpclett.1c01847} {\bibfield  {journal} {\bibinfo
   {journal} {J. Phys. Chem. Lett.}\ }\textbf {\bibinfo {volume} {12}},\
  \bibinfo {pages} {6974} (\bibinfo {year} {2021}{\natexlab{c}})}\BibitemShut
  {NoStop}%
\bibitem [{\citenamefont {Sch\"afer}(2022)}]{22Schafer}%
  \BibitemOpen
  \bibfield  {author} {\bibinfo {author} {\bibfnamefont {C.}~\bibnamefont
  {Sch\"afer}},\ }\href {https://doi.org/10.1021/acs.jpclett.2c01169}
  {\bibfield  {journal} {\bibinfo  {journal} {J. Phys. Chem. Lett.}\ }\textbf
  {\bibinfo {volume} {13}},\ \bibinfo {pages} {6905} (\bibinfo {year}
  {2022})}\BibitemShut {NoStop}%
\bibitem [{\citenamefont {Sch{\"a}fer}\ \emph {et~al.}(2022)\citenamefont
  {Sch{\"a}fer}, \citenamefont {Flick}, \citenamefont {Ronca}, \citenamefont
  {Narang},\ and\ \citenamefont {Rubio}}]{22ScFlRo}%
  \BibitemOpen
  \bibfield  {author} {\bibinfo {author} {\bibfnamefont {C.}~\bibnamefont
  {Sch{\"a}fer}}, \bibinfo {author} {\bibfnamefont {J.}~\bibnamefont {Flick}},
  \bibinfo {author} {\bibfnamefont {E.}~\bibnamefont {Ronca}}, \bibinfo
  {author} {\bibfnamefont {P.}~\bibnamefont {Narang}},\ and\ \bibinfo {author}
  {\bibfnamefont {A.}~\bibnamefont {Rubio}},\ }\href
  {https://doi.org/10.1038/s41467-022-35363-6} {\bibfield  {journal} {\bibinfo
  {journal} {Nat. Commun.}\ }\textbf {\bibinfo {volume} {13}},\ \bibinfo
  {pages} {7817} (\bibinfo {year} {2022})}\BibitemShut {NoStop}%
\bibitem [{\citenamefont {Pannir-Sivajothi}\ \emph {et~al.}(2022)\citenamefont
  {Pannir-Sivajothi}, \citenamefont {Campos-Gonzalez-Angulo}, \citenamefont
  {Mart{\'i}nez-Mart{\'i}nez}, \citenamefont {Sinha},\ and\ \citenamefont
  {Yuen-Zhou}}]{22PaCaMa}%
  \BibitemOpen
  \bibfield  {author} {\bibinfo {author} {\bibfnamefont {S.}~\bibnamefont
  {Pannir-Sivajothi}}, \bibinfo {author} {\bibfnamefont {J.~A.}\ \bibnamefont
  {Campos-Gonzalez-Angulo}}, \bibinfo {author} {\bibfnamefont {L.~A.}\
  \bibnamefont {Mart{\'i}nez-Mart{\'i}nez}}, \bibinfo {author} {\bibfnamefont
  {S.}~\bibnamefont {Sinha}},\ and\ \bibinfo {author} {\bibfnamefont
  {J.}~\bibnamefont {Yuen-Zhou}},\ }\href
  {https://doi.org/10.1038/s41467-022-29290-9} {\bibfield  {journal} {\bibinfo
  {journal} {Nat. Commun.}\ }\textbf {\bibinfo {volume} {13}},\ \bibinfo
  {pages} {1645} (\bibinfo {year} {2022})}\BibitemShut {NoStop}%
\bibitem [{\citenamefont {Du}\ and\ \citenamefont {Yuen-Zhou}(2022)}]{22DuYu}%
  \BibitemOpen
  \bibfield  {author} {\bibinfo {author} {\bibfnamefont {M.}~\bibnamefont
  {Du}}\ and\ \bibinfo {author} {\bibfnamefont {J.}~\bibnamefont {Yuen-Zhou}},\
  }\href {https://doi.org/10.1103/PhysRevLett.128.096001} {\bibfield  {journal}
  {\bibinfo  {journal} {Phys. Rev. Lett.}\ }\textbf {\bibinfo {volume} {128}},\
  \bibinfo {pages} {096001} (\bibinfo {year} {2022})}\BibitemShut {NoStop}%
\bibitem [{\citenamefont {Sun}\ and\ \citenamefont {Vendrell}(2022)}]{22SuVe}%
  \BibitemOpen
  \bibfield  {author} {\bibinfo {author} {\bibfnamefont {J.}~\bibnamefont
  {Sun}}\ and\ \bibinfo {author} {\bibfnamefont {O.}~\bibnamefont {Vendrell}},\
  }\href {https://doi.org/10.1021/acs.jpclett.2c00974} {\bibfield  {journal}
  {\bibinfo  {journal} {J. Phys. Chem. Lett.}\ }\textbf {\bibinfo {volume}
  {13}},\ \bibinfo {pages} {4441} (\bibinfo {year} {2022})}\BibitemShut
  {NoStop}%
\bibitem [{\citenamefont {Kowalewski}\ and\ \citenamefont
  {Mukamel}(2017)}]{17KoMu}%
  \BibitemOpen
  \bibfield  {author} {\bibinfo {author} {\bibfnamefont {M.}~\bibnamefont
  {Kowalewski}}\ and\ \bibinfo {author} {\bibfnamefont {S.}~\bibnamefont
  {Mukamel}},\ }\href {https://doi.org/10.1073/pnas.1702160114} {\bibfield
  {journal} {\bibinfo  {journal} {Proc. Natl. Acad. Sci. U.S.A.}\ }\textbf
  {\bibinfo {volume} {114}},\ \bibinfo {pages} {3278} (\bibinfo {year}
  {2017})}\BibitemShut {NoStop}%
\bibitem [{\citenamefont {Feist}\ \emph {et~al.}(2018)\citenamefont {Feist},
  \citenamefont {Galego},\ and\ \citenamefont {Garcia-Vidal}}]{18FeGaGa}%
  \BibitemOpen
  \bibfield  {author} {\bibinfo {author} {\bibfnamefont {J.}~\bibnamefont
  {Feist}}, \bibinfo {author} {\bibfnamefont {J.}~\bibnamefont {Galego}},\ and\
  \bibinfo {author} {\bibfnamefont {F.}~\bibnamefont {Garcia-Vidal}},\ }\href
  {https://doi.org/10.1021/acsphotonics.7b00680} {\bibfield  {journal}
  {\bibinfo  {journal} {ACS Photonics}\ }\textbf {\bibinfo {volume} {5}},\
  \bibinfo {pages} {205} (\bibinfo {year} {2018})}\BibitemShut {NoStop}%
\bibitem [{\citenamefont {Ribeiro}\ \emph {et~al.}(2018)\citenamefont
  {Ribeiro}, \citenamefont {Mart\'inez-Mart\'inez}, \citenamefont {Du},
  \citenamefont {Campos-Gonzalez-Angulo},\ and\ \citenamefont
  {Yuen-Zhou}}]{18RiMaDu}%
  \BibitemOpen
  \bibfield  {author} {\bibinfo {author} {\bibfnamefont {R.}~\bibnamefont
  {Ribeiro}}, \bibinfo {author} {\bibfnamefont {L.}~\bibnamefont
  {Mart\'inez-Mart\'inez}}, \bibinfo {author} {\bibfnamefont {M.}~\bibnamefont
  {Du}}, \bibinfo {author} {\bibfnamefont {J.}~\bibnamefont
  {Campos-Gonzalez-Angulo}},\ and\ \bibinfo {author} {\bibfnamefont
  {J.}~\bibnamefont {Yuen-Zhou}},\ }\href {https://doi.org/10.1039/c8sc01043a}
  {\bibfield  {journal} {\bibinfo  {journal} {Chem. Sci.}\ }\textbf {\bibinfo
  {volume} {9}},\ \bibinfo {pages} {6325} (\bibinfo {year} {2018})}\BibitemShut
  {NoStop}%
\bibitem [{\citenamefont {Herrera}\ and\ \citenamefont
  {Owrutsky}(2020)}]{20HeOw}%
  \BibitemOpen
  \bibfield  {author} {\bibinfo {author} {\bibfnamefont {F.}~\bibnamefont
  {Herrera}}\ and\ \bibinfo {author} {\bibfnamefont {J.}~\bibnamefont
  {Owrutsky}},\ }\href {https://doi.org/10.1063/1.5136320} {\bibfield
  {journal} {\bibinfo  {journal} {J. Chem. Phys.}\ }\textbf {\bibinfo {volume}
  {152}},\ \bibinfo {pages} {100902} (\bibinfo {year} {2020})}\BibitemShut
  {NoStop}%
\bibitem [{\citenamefont {Li}\ \emph {et~al.}(2022{\natexlab{b}})\citenamefont
  {Li}, \citenamefont {Cui}, \citenamefont {Subotnik},\ and\ \citenamefont
  {Nitzan}}]{22LiCuSu}%
  \BibitemOpen
  \bibfield  {author} {\bibinfo {author} {\bibfnamefont {T.~E.}\ \bibnamefont
  {Li}}, \bibinfo {author} {\bibfnamefont {B.}~\bibnamefont {Cui}}, \bibinfo
  {author} {\bibfnamefont {J.~E.}\ \bibnamefont {Subotnik}},\ and\ \bibinfo
  {author} {\bibfnamefont {A.}~\bibnamefont {Nitzan}},\ }\href
  {https://doi.org/10.1146/annurev-physchem-090519-042621} {\bibfield
  {journal} {\bibinfo  {journal} {Annu. Rev. Phys. Chem.}\ }\textbf {\bibinfo
  {volume} {73}},\ \bibinfo {pages} {43} (\bibinfo {year}
  {2022}{\natexlab{b}})}\BibitemShut {NoStop}%
\bibitem [{\citenamefont {Fregoni}\ \emph {et~al.}(2022)\citenamefont
  {Fregoni}, \citenamefont {Garcia-Vidal},\ and\ \citenamefont
  {Feist}}]{22FrGaFe}%
  \BibitemOpen
  \bibfield  {author} {\bibinfo {author} {\bibfnamefont {J.}~\bibnamefont
  {Fregoni}}, \bibinfo {author} {\bibfnamefont {F.~J.}\ \bibnamefont
  {Garcia-Vidal}},\ and\ \bibinfo {author} {\bibfnamefont {J.}~\bibnamefont
  {Feist}},\ }\href {https://doi.org/10.1021/acsphotonics.1c01749} {\bibfield
  {journal} {\bibinfo  {journal} {ACS Photonics}\ }\textbf {\bibinfo {volume}
  {9}},\ \bibinfo {pages} {1096} (\bibinfo {year} {2022})}\BibitemShut
  {NoStop}%
\bibitem [{\citenamefont {S\'anchez-Barquilla}\ \emph
  {et~al.}(2022)\citenamefont {S\'anchez-Barquilla}, \citenamefont
  {Fern\'andez-Dom\'inguez}, \citenamefont {Feist},\ and\ \citenamefont
  {Garc\'ia-Vidal}}]{22SaFeFe}%
  \BibitemOpen
  \bibfield  {author} {\bibinfo {author} {\bibfnamefont {M.}~\bibnamefont
  {S\'anchez-Barquilla}}, \bibinfo {author} {\bibfnamefont {A.~I.}\
  \bibnamefont {Fern\'andez-Dom\'inguez}}, \bibinfo {author} {\bibfnamefont
  {J.}~\bibnamefont {Feist}},\ and\ \bibinfo {author} {\bibfnamefont {F.~J.}\
  \bibnamefont {Garc\'ia-Vidal}},\ }\href
  {https://doi.org/10.1021/acsphotonics.2c00048} {\bibfield  {journal}
  {\bibinfo  {journal} {ACS Photonics}\ }\textbf {\bibinfo {volume} {9}},\
  \bibinfo {pages} {1830} (\bibinfo {year} {2022})}\BibitemShut {NoStop}%
\bibitem [{\citenamefont {Ruggenthaler}\ \emph {et~al.}(2014)\citenamefont
  {Ruggenthaler}, \citenamefont {Flick}, \citenamefont {Pellegrini},
  \citenamefont {Appel}, \citenamefont {Tokatly},\ and\ \citenamefont
  {Rubio}}]{14RuFlPe}%
  \BibitemOpen
  \bibfield  {author} {\bibinfo {author} {\bibfnamefont {M.}~\bibnamefont
  {Ruggenthaler}}, \bibinfo {author} {\bibfnamefont {J.}~\bibnamefont {Flick}},
  \bibinfo {author} {\bibfnamefont {C.}~\bibnamefont {Pellegrini}}, \bibinfo
  {author} {\bibfnamefont {H.}~\bibnamefont {Appel}}, \bibinfo {author}
  {\bibfnamefont {I.~V.}\ \bibnamefont {Tokatly}},\ and\ \bibinfo {author}
  {\bibfnamefont {A.}~\bibnamefont {Rubio}},\ }\href
  {https://doi.org/10.1103/PhysRevA.90.012508} {\bibfield  {journal} {\bibinfo
  {journal} {Phys. Rev. A}\ }\textbf {\bibinfo {volume} {90}},\ \bibinfo
  {pages} {012508} (\bibinfo {year} {2014})}\BibitemShut {NoStop}%
\bibitem [{\citenamefont {Flick}\ \emph {et~al.}(2017)\citenamefont {Flick},
  \citenamefont {Ruggenthaler}, \citenamefont {Appel},\ and\ \citenamefont
  {Rubio}}]{17FlRuAp}%
  \BibitemOpen
  \bibfield  {author} {\bibinfo {author} {\bibfnamefont {J.}~\bibnamefont
  {Flick}}, \bibinfo {author} {\bibfnamefont {M.}~\bibnamefont {Ruggenthaler}},
  \bibinfo {author} {\bibfnamefont {H.}~\bibnamefont {Appel}},\ and\ \bibinfo
  {author} {\bibfnamefont {A.}~\bibnamefont {Rubio}},\ }\href
  {https://doi.org/10.1073/pnas.1615509114} {\bibfield  {journal} {\bibinfo
  {journal} {Proc. Natl. Acad. Sci. U.S.A.}\ }\textbf {\bibinfo {volume}
  {114}},\ \bibinfo {pages} {3026} (\bibinfo {year} {2017})}\BibitemShut
  {NoStop}%
\bibitem [{\citenamefont {Haugland}\ \emph {et~al.}(2020)\citenamefont
  {Haugland}, \citenamefont {Ronca}, \citenamefont {Kj\o{}nstad}, \citenamefont
  {Rubio},\ and\ \citenamefont {Koch}}]{20HaRoKj}%
  \BibitemOpen
  \bibfield  {author} {\bibinfo {author} {\bibfnamefont {T.~S.}\ \bibnamefont
  {Haugland}}, \bibinfo {author} {\bibfnamefont {E.}~\bibnamefont {Ronca}},
  \bibinfo {author} {\bibfnamefont {E.~F.}\ \bibnamefont {Kj\o{}nstad}},
  \bibinfo {author} {\bibfnamefont {A.}~\bibnamefont {Rubio}},\ and\ \bibinfo
  {author} {\bibfnamefont {H.}~\bibnamefont {Koch}},\ }\href
  {https://doi.org/10.1103/PhysRevX.10.041043} {\bibfield  {journal} {\bibinfo
  {journal} {Phys. Rev. X}\ }\textbf {\bibinfo {volume} {10}},\ \bibinfo
  {pages} {041043} (\bibinfo {year} {2020})}\BibitemShut {NoStop}%
\bibitem [{\citenamefont {Sch\"{a}fer}\ \emph {et~al.}(2021)\citenamefont
  {Sch\"{a}fer}, \citenamefont {Buchholz}, \citenamefont {Penz}, \citenamefont
  {Ruggenthaler},\ and\ \citenamefont {Rubio}}]{21ScBuPe}%
  \BibitemOpen
  \bibfield  {author} {\bibinfo {author} {\bibfnamefont {C.}~\bibnamefont
  {Sch\"{a}fer}}, \bibinfo {author} {\bibfnamefont {F.}~\bibnamefont
  {Buchholz}}, \bibinfo {author} {\bibfnamefont {M.}~\bibnamefont {Penz}},
  \bibinfo {author} {\bibfnamefont {M.}~\bibnamefont {Ruggenthaler}},\ and\
  \bibinfo {author} {\bibfnamefont {A.}~\bibnamefont {Rubio}},\ }\href
  {https://doi.org/10.1073/pnas.2110464118} {\bibfield  {journal} {\bibinfo
  {journal} {Proc. Natl. Acad. Sci. U.S.A.}\ }\textbf {\bibinfo {volume}
  {118}},\ \bibinfo {pages} {e2110464118} (\bibinfo {year} {2021})}\BibitemShut
  {NoStop}%
\bibitem [{\citenamefont {Sch\"afer}\ and\ \citenamefont
  {Johansson}(2022)}]{22ScJo}%
  \BibitemOpen
  \bibfield  {author} {\bibinfo {author} {\bibfnamefont {C.}~\bibnamefont
  {Sch\"afer}}\ and\ \bibinfo {author} {\bibfnamefont {G.}~\bibnamefont
  {Johansson}},\ }\href {https://doi.org/10.1103/PhysRevLett.128.156402}
  {\bibfield  {journal} {\bibinfo  {journal} {Phys. Rev. Lett.}\ }\textbf
  {\bibinfo {volume} {128}},\ \bibinfo {pages} {156402} (\bibinfo {year}
  {2022})}\BibitemShut {NoStop}%
\bibitem [{\citenamefont {Jaynes}\ and\ \citenamefont
  {Cummings}(1963)}]{63JaCu}%
  \BibitemOpen
  \bibfield  {author} {\bibinfo {author} {\bibfnamefont {E.}~\bibnamefont
  {Jaynes}}\ and\ \bibinfo {author} {\bibfnamefont {F.}~\bibnamefont
  {Cummings}},\ }\href {https://doi.org/10.1109/PROC.1963.1664} {\bibfield
  {journal} {\bibinfo  {journal} {Proc. IEEE}\ }\textbf {\bibinfo {volume}
  {51}},\ \bibinfo {pages} {89} (\bibinfo {year} {1963})}\BibitemShut {NoStop}%
\bibitem [{\citenamefont {Tavis}\ and\ \citenamefont
  {Cummings}(1968)}]{68TaCu}%
  \BibitemOpen
  \bibfield  {author} {\bibinfo {author} {\bibfnamefont {M.}~\bibnamefont
  {Tavis}}\ and\ \bibinfo {author} {\bibfnamefont {F.~W.}\ \bibnamefont
  {Cummings}},\ }\href {https://doi.org/10.1103/PhysRev.170.379} {\bibfield
  {journal} {\bibinfo  {journal} {Phys. Rev.}\ }\textbf {\bibinfo {volume}
  {170}},\ \bibinfo {pages} {379} (\bibinfo {year} {1968})}\BibitemShut
  {NoStop}%
\bibitem [{\citenamefont {Hepp}\ and\ \citenamefont {Lieb}(1973)}]{73HeLi}%
  \BibitemOpen
  \bibfield  {author} {\bibinfo {author} {\bibfnamefont {K.}~\bibnamefont
  {Hepp}}\ and\ \bibinfo {author} {\bibfnamefont {E.~H.}\ \bibnamefont
  {Lieb}},\ }\href
  {https://doi.org/https://doi.org/10.1016/0003-4916(73)90039-0} {\bibfield
  {journal} {\bibinfo  {journal} {Ann. Phys.}\ }\textbf {\bibinfo {volume}
  {76}},\ \bibinfo {pages} {360} (\bibinfo {year} {1973})}\BibitemShut
  {NoStop}%
\bibitem [{\citenamefont {Kardar}(2007)}]{07Kardar}%
  \BibitemOpen
  \bibfield  {author} {\bibinfo {author} {\bibfnamefont {M.}~\bibnamefont
  {Kardar}},\ }\href {https://doi.org/10.1017/CBO9780511815898} {\emph
  {\bibinfo {title} {Statistical Physics of Particles}}}\ (\bibinfo
  {publisher} {Cambridge University Press},\ \bibinfo {address} {Cambridge},\
  \bibinfo {year} {2007})\BibitemShut {NoStop}%
\bibitem [{\citenamefont {Pilar}\ \emph {et~al.}(2020)\citenamefont {Pilar},
  \citenamefont {De~Bernardis},\ and\ \citenamefont {Rabl}}]{20PiBeRa}%
  \BibitemOpen
  \bibfield  {author} {\bibinfo {author} {\bibfnamefont {P.}~\bibnamefont
  {Pilar}}, \bibinfo {author} {\bibfnamefont {D.}~\bibnamefont
  {De~Bernardis}},\ and\ \bibinfo {author} {\bibfnamefont {P.}~\bibnamefont
  {Rabl}},\ }\href {https://doi.org/10.22331/q-2020-09-28-335} {\bibfield
  {journal} {\bibinfo  {journal} {{Quantum}}\ }\textbf {\bibinfo {volume}
  {4}},\ \bibinfo {pages} {335} (\bibinfo {year} {2020})}\BibitemShut {NoStop}%
\bibitem [{\citenamefont {Cs\'asz\'ar}\ \emph {et~al.}(2012)\citenamefont
  {Cs\'asz\'ar}, \citenamefont {F\'abri}, \citenamefont {Szidarovszky},
  \citenamefont {M\'atyus}, \citenamefont {Furtenbacher},\ and\ \citenamefont
  {Czak\'o}}]{12CsFaSz}%
  \BibitemOpen
  \bibfield  {author} {\bibinfo {author} {\bibfnamefont {A.~G.}\ \bibnamefont
  {Cs\'asz\'ar}}, \bibinfo {author} {\bibfnamefont {C.}~\bibnamefont
  {F\'abri}}, \bibinfo {author} {\bibfnamefont {T.}~\bibnamefont
  {Szidarovszky}}, \bibinfo {author} {\bibfnamefont {E.}~\bibnamefont
  {M\'atyus}}, \bibinfo {author} {\bibfnamefont {T.}~\bibnamefont
  {Furtenbacher}},\ and\ \bibinfo {author} {\bibfnamefont {G.}~\bibnamefont
  {Czak\'o}},\ }\href {https://doi.org/10.1039/C1CP21830A} {\bibfield
  {journal} {\bibinfo  {journal} {Phys. Chem. Chem. Phys.}\ }\textbf {\bibinfo
  {volume} {14}},\ \bibinfo {pages} {1085} (\bibinfo {year}
  {2012})}\BibitemShut {NoStop}%
\bibitem [{\citenamefont {Cohen-Tannoudji}\ \emph {et~al.}(2004)\citenamefont
  {Cohen-Tannoudji}, \citenamefont {Dupont-Roc},\ and\ \citenamefont
  {Grynberg}}]{04CoDuGr}%
  \BibitemOpen
  \bibfield  {author} {\bibinfo {author} {\bibfnamefont {C.}~\bibnamefont
  {Cohen-Tannoudji}}, \bibinfo {author} {\bibfnamefont {J.}~\bibnamefont
  {Dupont-Roc}},\ and\ \bibinfo {author} {\bibfnamefont {G.}~\bibnamefont
  {Grynberg}},\ }\href@noop {} {\emph {\bibinfo {title} {Atom-Photon
  Interactions: Basic Processes and Applications}}}\ (\bibinfo  {publisher}
  {Wiley-VCH Verlag GmbH and Co. KGaA},\ \bibinfo {address} {Weinheim},\
  \bibinfo {year} {2004})\BibitemShut {NoStop}%
\bibitem [{\citenamefont {Sidler}\ \emph {et~al.}(2023)\citenamefont {Sidler},
  \citenamefont {Schnappinger}, \citenamefont {Obzhirov}, \citenamefont
  {Ruggenthaler}, \citenamefont {Kowalewski},\ and\ \citenamefont
  {Rubio}}]{23SiScOb}%
  \BibitemOpen
  \bibfield  {author} {\bibinfo {author} {\bibfnamefont {D.}~\bibnamefont
  {Sidler}}, \bibinfo {author} {\bibfnamefont {T.}~\bibnamefont
  {Schnappinger}}, \bibinfo {author} {\bibfnamefont {A.}~\bibnamefont
  {Obzhirov}}, \bibinfo {author} {\bibfnamefont {M.}~\bibnamefont
  {Ruggenthaler}}, \bibinfo {author} {\bibfnamefont {M.}~\bibnamefont
  {Kowalewski}},\ and\ \bibinfo {author} {\bibfnamefont {A.}~\bibnamefont
  {Rubio}},\ }\href {https://doi.org/10.48550/arXiv.2306.06004} {\bibinfo
  {title} {Unraveling a cavity induced molecular polarization mechanism from
  collective vibrational strong coupling}} (\bibinfo {year} {2023}),\ \Eprint
  {https://arxiv.org/abs/2306.06004} {arXiv:2306.06004 [quant-ph]} \BibitemShut
  {NoStop}%
\bibitem [{\citenamefont {Cohen-Tannoudji}\ \emph {et~al.}(1977)\citenamefont
  {Cohen-Tannoudji}, \citenamefont {Diu},\ and\ \citenamefont
  {Lalo{\"e}}}]{77CoDiLa}%
  \BibitemOpen
  \bibfield  {author} {\bibinfo {author} {\bibfnamefont {C.}~\bibnamefont
  {Cohen-Tannoudji}}, \bibinfo {author} {\bibfnamefont {B.}~\bibnamefont
  {Diu}},\ and\ \bibinfo {author} {\bibfnamefont {F.}~\bibnamefont
  {Lalo{\"e}}},\ }\href {https://cds.cern.ch/record/101367} {\emph {\bibinfo
  {title} {{Quantum mechanics; 1st ed.}}}}\ (\bibinfo  {publisher} {Wiley},\
  \bibinfo {address} {New York, NY},\ \bibinfo {year} {1977})\BibitemShut
  {NoStop}%
\end{thebibliography}%

\clearpage

\begin{center}
{\huge Supplementary Material}
\end{center}

\section*{Theoretical derivations}
In what follows, two extensive theoretical derivations are presented, namely, a step-by-step verification of Eqs. (16) and (18) of the manuscript is described in detail.
Let us start with proving the formula for $\textrm{tr}[\hat{\Delta}_1]$ (see Eq. (16) of the manuscript),
\begin{gather}
    \textrm{tr}[\hat{\Delta}_1] = 
    \sum_{\alpha n} \langle \alpha n \big| \sum_{k=1}^\infty \frac{(-\beta)^k}{k!} \sum_{i=0}^{k-1} \hat{H}_0^i ~ \Delta\hat{H} ~ \hat{H}_0^{k-(i+1)} \big| \alpha n \rangle \label{eq:trd1} \\ \nonumber
    = \sum_{\alpha n} \sum_{k=1}^\infty \frac{(-\beta)^k}{k!} \sum_{i=0}^{k-1} E_{\alpha n}^i \Delta H_{\alpha n, \alpha n} E_{\alpha n}^{k-(i+1)} 
    = \sum_{\alpha n} \sum_{k=1}^\infty \frac{(-\beta)^{k}}{k!} k E_{\alpha n}^{k-1} \Delta H_{\alpha n, \alpha n} \\ \nonumber
    = -\beta \sum_{\alpha n} \exp (-\beta E_{\alpha n}) \Delta H_{\alpha n, \alpha n} 
    = -\frac{\beta g^2}{\hbar \omega_\textrm{c}} \sum_{\alpha} \exp (-\beta E_{\alpha}) (\mu^2)_{\alpha\alpha} \sum_{n=0}^\infty \exp (-\beta \epsilon_n) \\ \nonumber
    = -\frac{\beta g^2 Q_\textrm{c}}{\hbar \omega_\textrm{c}} \sum_{\alpha} \exp (-\beta E_{\alpha}) (\mu^2)_{\alpha\alpha}.
\end{gather}
A few comments are necessary: 
(i) in the first two lines the relation $\hat{H}_0 | \alpha n \rangle = E_{\alpha n} | \alpha n \rangle$ is used
(see also Eq. (8) of the manuscript),
(ii) $E_{\alpha n} = E_{\alpha} + \epsilon_n$ and $\Delta H_{\alpha n, \alpha n} = \langle \alpha n \big| \Delta\hat{H} | \alpha n \rangle$,
(iii) the sum over $i$ in the second line gives a factor of $k$,
(iv) the infinite sum over $k$ in the second line is evaluated through the relation
\begin{equation}
    \sum_{k=1}^\infty \frac{(-\beta)^k}{k!} k E_{\alpha n}^{k-1} = -\beta \sum_{k=1}^\infty \frac{(-\beta E_{\alpha n})^{k-1}}{(k-1)!} = 
    	-\beta \sum_{k=0}^\infty \frac{(-\beta E_{\alpha n})^k}{k!} = -\beta \exp(-\beta E_{\alpha n})
\end{equation}
where an index shift in $k$ is applied,
(v) $Q_\textrm{c} = \sum_{n=0}^\infty \exp (-\beta \epsilon_n)$ is the partition function of the cavity mode.
As diagonal matrix elements of the cavity coordinate $q_\textrm{c}$ are zero ($\langle n | q_\textrm{c} | n \rangle = 0$),
only the second term of $\Delta\hat{H}$ (see Eq. (4) of the manuscript) gives nonvanishing diagonal matrix elements,
therefore, $\Delta H_{\alpha n, \alpha n} = \frac{g^2}{\hbar \omega_\textrm{c}} (\mu^2)_{\alpha\alpha}$.
Thus, no terms linear in $\mu$ appear in $\textrm{tr}[\hat{\Delta}_1]$.

The derivation of Eq. (18) of the manuscript is performed in two steps. First, we consider
\begin{gather}
    \textrm{tr}[\hat{\Delta}_2] = 
    \sum_{\alpha n} \langle \alpha n \big| \sum_{k=2}^\infty \frac{(-\beta)^k}{k!} \sum_{i=0}^{k-2} ~ \sum_{j=0}^{k-(i+2)} \hat{H}_0^i ~ \Delta\hat{H} ~ \hat{H}_0^j ~ 
        \Delta\hat{H} ~ \hat{H}_0^{k-(i+j+2)} \big| \alpha n \rangle \label{eq:trd21} \\ \nonumber
    = \sum_{\alpha n} \sum_{\alpha' n'} \sum_{k=2}^\infty \frac{(-\beta)^k}{k!} \sum_{i=0}^{k-2} ~ \sum_{j=0}^{k-(i+2)} 
        \langle \alpha n \big| \hat{H}_0^i ~ \Delta\hat{H} ~ \hat{H}_0^j \big| \alpha' n' \rangle 
        \langle \alpha' n' \big| \Delta\hat{H} ~ \hat{H}_0^{k-(i+j+2)} \big| \alpha n  \rangle \\ \nonumber
    = \sum_{\alpha n} \sum_{\alpha' n'} \sum_{k=2}^\infty \frac{(-\beta)^k}{k!} \sum_{i=0}^{k-2} ~ \sum_{j=0}^{k-(i+2)} E_{\alpha n}^i E_{\alpha' n'}^j
        E_{\alpha n}^{k-(i+j+2)} |\Delta H_{\alpha n, \alpha' n'}|^2 \\ \nonumber
    = \sum_{\alpha n} \sum_{\alpha' n'} 
        \frac{\exp(-\beta E_{\alpha' n'}) - \exp(-\beta E_{\alpha n}) - \beta (E_{\alpha n}-E_{\alpha' n'}) \exp(-\beta E_{\alpha n})}{(E_{\alpha n}-E_{\alpha' n'})^2} 
        |\Delta H_{\alpha n, \alpha' n'}|^2 \\ \nonumber
    = \beta \sum_{\alpha n} \sum_{\alpha' n'} \frac{\exp(-\beta E_{\alpha n})}{E_{\alpha' n'}-E_{\alpha n}} |\Delta H_{\alpha n, \alpha' n'}|^2 = 
    \frac{2\omega_\textrm{c} \beta g^2}{\hbar} \sum_{\alpha n} \sum_{\alpha' n'} \frac{\exp(-\beta E_{\alpha n})}{E_{\alpha' n'}-E_{\alpha n}} |\mu_{\alpha\alpha'} (q_\textrm{c})_{nn'}|^2
\end{gather}
where
(i) the comments below Eq. \eqref{eq:trd1} apply to the current derivation as well,
(ii) a resolution of identity is inserted between $\hat{H}_0^j$ and $\Delta\hat{H}$ in the second line,
(iii) the double sum over $i$ and $j$ can be evaluated analytically (see the next paragraph for more explanation),
(iv) terms that change sign under the transformation $\alpha n \leftrightarrow \alpha' n'$ in the summation over $\alpha n$ and $\alpha' n'$ sum up to zero
(see fourth and fifth lines),
(v) since we keep terms up to second order in $\mu$, the approximation 
$\Delta H_{\alpha n, \alpha' n'} \approx -g \sqrt{\frac{2 \omega_\textrm{c}}{\hbar}} \mu_{\alpha\alpha'} (q_\textrm{c})_{nn'}$ is made in the last line.

Next, starting from the partial sum formula for the geometric series, that is,
\begin{equation}
    \sum_{k=0}^n q^k = \frac{q^{n+1}-1}{q-1},
\label{eq:psum}
\end{equation}
the double sum over $i$ and $j$ in Eq. \eqref{eq:trd21} is evaluated. With the partial sum formula of Eq. \eqref{eq:psum}, one can easily prove that
\begin{equation}
    \sum_{i=0}^{k-2} ~ \sum_{j=0}^{k-(i+2)} q^j = \sum_{i=0}^{k-2} \frac{q^{k-i-1}-1}{q-1} = \frac{q^{k-1}}{q-1} \sum_{i=0}^{k-2} \left( \frac{1}{q} \right)^i - \frac{k-1}{q-1} =
        \frac{q^k+k(1-q)-1}{(q-1)^2}.
\label{eq:ds}
\end{equation}
Using the simplified notations $a = E_{\alpha' n'}$ and $b = E_{\alpha n}$ the double sum over $i$ and $j$ in Eq. \eqref{eq:trd21} becomes
\begin{gather}
    \sum_{k=2}^\infty \frac{(-\beta)^k}{k!} \sum_{i=0}^{k-2} ~ \sum_{j=0}^{k-(i+2)} b^i a^j b^{k-(i+j+2)} =
         \sum_{k=2}^\infty \frac{(-\beta)^k}{k!} b^{k-2} \sum_{i=0}^{k-2} ~ \sum_{j=0}^{k-(i+2)} \left( \frac{a}{b} \right)^j \\ \nonumber
    = \sum_{k=2}^\infty \frac{(-\beta)^k}{k!} b^{k-2} \frac{\left( \frac{a}{b} \right)^k+k \left(1-\frac{a}{b} \right)-1}{(\frac{a}{b}-1)^2} =
        \frac{1}{(a-b)^2} \sum_{k=2}^\infty \frac{(-\beta)^k}{k!} a^k \\ \nonumber
    + \frac{1}{1-\frac{a}{b}} \sum_{k=2}^\infty \frac{(-\beta)^k}{k!} k b^{k-2} - \frac{1}{\left( 1-\frac{a}{b} \right)^2} \sum_{k=2}^\infty \frac{(-\beta)^k}{k!} b^{k-2} \\ \nonumber
    = \frac{\exp(-\beta a) - \exp(-\beta b) - \beta (b-a) \exp(-\beta b)}{(b-a)^2}
\end{gather}
where $a/b$ plays the role of $q$ in Eq. \eqref{eq:ds}.

The last step to complete the proof requires the relations
\begin{equation}
    (q_\textrm{c})_{nn'} = \langle n | q_\textrm{c} | n' \rangle = \sqrt{\frac{\hbar}{2\omega_\textrm{c}}} \left( \sqrt{n} \delta_{n,n'+1} + \sqrt{n+1} \delta_{n,n'-1} \right)
\label{eq:qcMx}
\end{equation}
and
\begin{gather}
    \sum_{n=0}^\infty (n+1) \exp(-\beta \epsilon_n) = \exp(\beta \hbar \omega_\textrm{c}/2) Q_\textrm{c}^2 \\ \nonumber
    \sum_{n=0}^\infty (n+1) \exp(-\beta \epsilon_{n+1}) = \exp(-\beta \hbar \omega_\textrm{c}/2) Q_\textrm{c}^2 
\end{gather}
which can be verified easily if one considers the energy levels $\epsilon_n = \hbar \omega_\textrm{c} \left( n+\frac{1}{2} \right)$
and the canonical partition function of the cavity mode,
$Q_\textrm{c} = \sum_{n=0}^\infty \exp (-\beta \epsilon_n) = \frac{\exp(-\beta \hbar \omega_\textrm{c}/2)}{1-\exp (-\beta \hbar \omega_\textrm{c})}$.
Thus, continuing Eq. \eqref{eq:trd21} yields
\begin{gather}
    \textrm{tr}[\hat{\Delta}_2] = 
    \frac{2\omega_\textrm{c} \beta g^2}{\hbar} \sum_{\alpha\alpha'} \exp(-\beta E_{\alpha}) |\mu_{\alpha\alpha'}|^2 
        \sum_{n=0}^\infty \sum_{n'=0}^\infty \frac{\exp(-\beta \epsilon_n)}{E_{\alpha'}-E_{\alpha}+\epsilon_{n'}-\epsilon_{n}}  |(q_\textrm{c})_{nn'}|^2 \label{eq:trd22} \\ \nonumber
    = \beta g^2 \sum_{\alpha\alpha'} \exp(-\beta E_{\alpha}) |\mu_{\alpha\alpha'}|^2
        \sum_{m=0}^\infty (m+1) \left( \frac{\exp(-\beta \epsilon_m)}{E_{\alpha'}-E_{\alpha} + \hbar \omega_\textrm{c}} +
            \frac{\exp(-\beta \epsilon_{m+1})}{E_{\alpha'}-E_{\alpha} - \hbar \omega_\textrm{c}} \right) \\ \nonumber
    = \beta g^2 Q_\textrm{c}^2 \sum_{\alpha\alpha'} \exp(-\beta E_{\alpha}) |\mu_{\alpha\alpha'}|^2
        \left( \frac{\exp(\beta \hbar \omega_\textrm{c}/2)}{E_{\alpha'}-E_{\alpha} + \hbar \omega_\textrm{c}} +
            \frac{\exp(-\beta \hbar \omega_\textrm{c}/2)}{E_{\alpha'}-E_{\alpha} - \hbar \omega_\textrm{c}} \right)
\end{gather}
where the double sum over $n$ and $n'$ in the first line can be rearranged to the sum over $m$ in the second line, which is explained by Table \ref{tbl:sum}.

\begin{table}
\caption{Tabular representation of the rearrengement of the double sum over $n$ and $n'$ to the sum over $m$ in Eq. \eqref{eq:trd22}.
Each line corresponds to a nonzero coordinate matrix element $(q_\textrm{c})_{nn'}$ (see Eq. \eqref{eq:qcMx}).
Note that each value of $m$ is equivalent to the index pairs $(n,n') = (m,m+1)$ and $(n,n') = (m+1,m)$.}
\label{tbl:sum}
\begin{center}
\begin{tabular}{ c|c|c|c } 
 $m$ & $n$ & $n'$ & $(q_\textrm{c})_{nn'} / \sqrt{\frac{\hbar}{2\omega_\textrm{c}}}$ \\
 \hline
 $0$ & $m=0$ & $m+1=1$ & $\sqrt{m+1} = 1$ \\
 $0$ & $m+1=1$ & $m=0$ & $\sqrt{m+1} = 1$ \\
 \hline
 $1$ & $m=1$ & $m+1=2$ & $\sqrt{m+1} = \sqrt{2}$ \\
 $1$ & $m+1=2$ & $m=1$ & $\sqrt{m+1} = \sqrt{2}$ \\
 \hline
 $2$ & $m=2$ & $m+1=3$ & $\sqrt{m+1} = \sqrt{3}$ \\
 $2$ & $m+1=3$ & $m=2$ & $\sqrt{m+1} = \sqrt{3}$ \\ 
 \hline
 \vdots & \vdots & \vdots & \vdots
\end{tabular}
\end{center}
\end{table}

\end{document}